\documentclass[journal]{IEEEtran}

\usepackage{enumitem}
\usepackage[yyyymmdd,hhmmss]{datetime}
\usepackage{amsmath, amsfonts, bm, amssymb, amsthm, comment}
\usepackage{array}
\usepackage{textcomp, mdframed}
\usepackage{stfloats}
\usepackage{verbatim}
\usepackage{graphicx, multirow}
\usepackage{cite}
\usepackage{kotex, lineno}
\usepackage[dvipsnames]{xcolor}
\usepackage{epsfig, epstopdf}
\usepackage{subfigure}
\usepackage{ulem}
\usepackage{soul}
\usepackage{algorithm, algorithmicx}
\usepackage[noEnd=false]{algpseudocodex}

\newtheorem{lemma}{Lemma}
\newtheorem{observation}{Observation}
\theoremstyle{definition}
\newtheorem{remark}{Remark}

\newcommand{\problem}[1]{\textsf{(P#1)}}
\renewcommand{\Re}[1]{\mathtt{Re}\{#1\}}
\renewcommand{\Im}[1]{\mathtt{Im}\{#1\}}
\newcommand{\tsuml}{\textstyle\sum\limits}

\newcommand{\NL}{\\}
\newcommand{\NNL}{\nonumber\\}
\newcommand{\NN}{\nonumber}

\newcommand{\Red}{\color{red}}

\newcommand{\blue}[1]{{\color{black}#1}}
\newcommand{\Blue}{\color{black}}
\newcommand{\Black}{\color{black}}

\newcommand{\magenta}[1]{{\color{black}#1}}

\newcommand{\dg}{\dagger}
\newcommand{\A}{\mathcal{A}}
\newcommand{\Ao}{\mathcal{A}^{\circ}}
\newcommand{\B}{\mathcal{B}}
\newcommand{\U}{\mathcal{U}}
\newcommand{\Uo}{\mathcal{U}^{\circ}}
\newcommand{\D}{\mathcal{D}}
\newcommand{\I}{\mathcal{I}}

\newcommand{\R}{\mathcal{R}}
\newcommand{\CB}{\mathcal{C}}

\renewcommand{\a}{\mathbf{a}}
\newcommand{\h}{\mathbf{h}}
\newcommand{\hh}{\hat{\mathbf{h}}}

\newcommand{\hg}{\hat{g}}

\newcommand{\w}{\mathbf{w}}

\newcommand{\x}{\mathbf{x}}

\newcommand{\e}{\mathbf{e}}

\newcommand{\E}{\mathbb{E}}
\newcommand{\C}{\mathbb{C}}

\newcommand{\KJcancel}[1]{}
\newcommand{\KJcancelc}[1]{{\color{red}#1}}

\newcommand{\KJcancelEQ}[1]{}
\renewcommand{\KJcancelEQ}[1]{{\color{red}#1}}

\newcommand{\teq}{\triangleq}

\newcommand{\JHbox}[1]{\vspace{.2in}\noindent\fbox{\begin{minipage}{\columnwidth}#1\end{minipage}}\vspace{.2in}}
\renewcommand{\JHbox}[1]{}

\newcommand{\PAP}[1]{\problem{#1{-\scriptsize\textsf{AP}}}}
\newcommand{\PUE}[1]{\problem{#1{-\scriptsize\textsf{UE} $k$}}}
\renewcommand{\PAP}[1]{\problem{$^{\scriptstyle\texttt{AP}}$#1}}
\renewcommand{\PUE}[1]{\problem{$^{\scriptstyle\texttt{UE}}$#1-$k$}}
\renewcommand{\PAP}[1]{\problem{$_{\scriptstyle\texttt{AP}}#1$}}
\renewcommand{\PUE}[1]{\problem{$_{\scriptstyle\texttt{UE}}$#1-$k$}}
\renewcommand{\PAP}[1]{{\normalfont\problem{$_{\scriptstyle\texttt{AP}}$#1}}}
\renewcommand{\PUE}[1]{{\normalfont\problem{$_{\scriptstyle\texttt{UE}}$#1-$k$}}}

\newcommand{\mk}{_{m,k}}

\newcommand{\mkl}{_{m,k,l}}
\renewcommand{\mkl}{_{m,k}^{\scriptscriptstyle(l)}}

\newcommand{\mkL}{_{m,k}^{\scriptscriptstyle(L)}}

\newcommand{\mkI}{_{m,k}^{\scriptscriptstyle(1)}}

\newcommand{\pmk}{{\red{p_{m,k}}}}

\renewcommand{\pmk}{{\red{p_m}}}

\renewcommand{\pmk}{{{p_m}}}

\newcommand{\Pmk}{{\red{(\vert\U_m\vert -1)\pmk}}}

\renewcommand{\Pmk}{{\red{\rho_m}}}

\renewcommand{\Pmk}{{{\rho_m}}}

\newcommand{\mkls}{_{m,k}^{{\scriptscriptstyle(l)}\star}}

\newcommand{\SGI}{{\scriptscriptstyle\mathsf{SGI}}}
\newcommand{\SGIDA}{{\SGI\downarrow}}
\newcommand{\lbound}{\texttt{lower}}

\newcommand{\IGI}{{\scriptscriptstyle\mathsf{IGI}}}

\newcommand{\CN}{\mathcal{CN}}
\newcommand{\Rsum}{\R_\textrm{\normalfont sum}}
\newcommand{\zero}{\mathbf{0}}

\renewcommand{\Delta}{\delta}

\definecolor{JHBGcolor}{HTML}{C7D7C7}

\sethlcolor{yellow}

\begin{document}
\title{User-Centric Association and Feedback Bit Allocation for FDD Cell-Free Massive MIMO}

\author{Kwangjae Lee,~\IEEEmembership{Graduate Student Member,~IEEE,} Jung Hoon Lee,~\IEEEmembership{Member,~IEEE}, and Wan Choi,~\IEEEmembership{Fellow,~IEEE}

\thanks{K. Lee and W. Choi are with the Department of Electrical and Computer Engineering, and the Institute of New Media and Communications, Seoul National University (SNU), Seoul 08826, Korea (e-mail: kwangjae@snu.ac.kr; wanchoi@snu.ac.kr).}
\thanks{J. H. Lee is with the Department of Electronics Engineering and Applied Communications Research Center, Hankuk University of Foreign Studies (HUFS), Yongin 17035, South Korea (e-mail: tantheta@hufs.ac.kr).}}

\maketitle

\begin{abstract}
In this paper, we introduce a novel approach to user-centric association and feedback bit allocation for the downlink of a cell-free massive MIMO (CF-mMIMO) system, operating under limited feedback constraints. In CF-mMIMO systems employing frequency division duplexing, each access point (AP) relies on channel information provided by its associated user equipments (UEs) for beamforming design. Since the uplink control channel is typically shared among UEs, we take account of each AP's total feedback budget, which is distributed among its associated UEs. By employing the Saleh-Valenzuela multi-resolvable path channel model with different average path gains, we first identify necessary feedback information for each UE, along with an appropriate codebook structure. This structure facilitates adaptive quantization of multiple paths based on their dominance. We then formulate a joint optimization problem addressing user-centric UE-AP association and feedback bit allocation. To address this challenge, we analyze the impact of feedback bit allocation and derive our proposed scheme from the solution of an alternative optimization problem aimed at devising long-term policies, explicitly considering the effects of feedback bit allocation. Numerical results show that our proposed scheme effectively enhances the performance of conventional approaches in CF-mMIMO systems.
\end{abstract}

\begin{IEEEkeywords}
Cell-free massive multiple-input multiple-output, frequency division duplexing, limited feedback, user-centric association.
\end{IEEEkeywords}

\section{Introduction}
\IEEEPARstart{T}{he} increasing demand for higher spectral efficiency is driving the advancement of 6G communication systems, which aim to surpass the capabilities of 5G communications. Cisco predicts that by 2029, the volume of data transmitted over networks will triple compared to 2023 \cite{Cisco}. Consequently, novel emerging technologies or substantial enhancements in existing ones are imperative to accommodate this surge in data. The main focuses of research in 5G technology include enhancing network density, harnessing millimeter-wave, and developing extensive multiple-input multiple-output (MIMO) systems \cite{Andrews2014}. While these technologies are experiencing gradual enhancements, they are anticipated to undergo significant evolution within the realm of 6G \cite{Akyildiz2020}.

One strategy for enhancing spectral efficiency involves employing macro-diversity through ultra-densification, achieved by deploying a greater number of access points (APs) within a network area. For the densification, heterogeneous networks were considered in 4G LTE communication systems \cite{Hoydis2011, Dhillon2012, Jo2012}, where small cells (pico-cells and femto-cells) are deployed along with existing macro-cells. This evolution progresses toward cell-free massive MIMO (CF-mMIMO) systems to enhance performance and ensure user fairness \cite{Ngo2017}. CF-mMIMO systems integrate various types of distributed AP systems such as distributed antenna systems, coordinated multi-point transmission, virtual cellular networks, and network MIMO \cite{Choi2007, Lee2016}. In these systems, all or a subset of APs serve user equipments (UEs), accommodating a wide range of scenarios and focusing on user-centric clustering.

When serving multiple UEs simultaneously, each AP requires channel state information (CSI) to mitigate inter-UE interference. Generally, acquiring CSI at a transmitter is deemed more straightforward in time division duplexing (TDD) systems compared to frequency division duplexing (FDD) ones. This is attributed to the reciprocity of uplink and downlink channels in TDD systems, enabling the transmitter to directly glean CSI from the uplink channel. Conversely, in FDD systems, where uplink and downlink channels operate independently, the transmitter relies on the receiver for quantized information. It was shown in \cite{Jindal2006} that the feedback overhead should be proportional to the transmit power in decibels to maintain a certain multiplexing gain in multiuser MIMO broadcast channels. Consequently, this has driven 5G/6G systems, particularly those employing mMIMO and CF-mMIMO, to adopt TDD operation modes \cite{Lu2014}.

However, TDD systems still grapple with numerous challenges in achieving precise channel reciprocity, such as calibration errors in RF chains, prompting a reevaluation of FDD-based 5G/6G communication systems. Many studies investigated the impact of channel quantization errors on various FDD wireless communication systems \cite{Love2005, Lee2012, Lee2013, Lee2015}. Especially in multiuser MIMO scenarios, channel quantization errors are crucial as they harm the multiplexing gain \cite{Jindal2006}. In the context of CF-mMIMO, efficient management of feedback overhead facilitates the exploitation of promising FDD CF-mMIMO systems that remain backward compatible with preceding FDD wireless communication systems.

Numerous studies have explored UE-AP association for various wireless communication systems \cite{Yoo2006, Yoo2007, Buzzi2017, Buzzi2020, Bjornson2020, Kim2020, Kim2021}. Traditionally, the association between UEs and APs is determined by the APs, relying on available information \cite{Yoo2006, Yoo2007, Buzzi2017, Buzzi2020}. However, CF-mMIMO systems exhibit a shift towards user-centric clustering, where each UE initiates clustering APs for association. In \cite{Bjornson2020}, the authors introduced dynamic cooperation clustering in TDD CF-mMIMO systems for user-centric UE-AP association. Additionally, in FDD CF-mMIMO systems, the authors of \cite{Kim2020, Kim2021} studied user-centric UE-AP association. Their approach sets the association first, independent of the feedback protocol, and then generates the feedback information. Nevertheless, there has been limited research on conducting user-centric UE-AP association in FDD CF-mMIMO systems, particularly considering the impact of imperfect CSI on system performance.

The literature concerning FDD CF-mMIMO systems predominantly focuses on multi-resolvable path channel models, such as ray-based and multipath Rician fading channel models \cite{Kim2020, Kim2021, Abdallah2020, Lee2022}. In these channel models, each channel comprises multiple resolvable paths (i.e., rays), exhibiting reciprocity between the angle of arrival (AoA) and the angle of departure (AoD) for each path. In \cite{Abdallah2020}, the authors explored transmit beamforming and receive combining based on angle information within the ray-based channel model, without any CSI feedback. Subsequent works by the authors in \cite{Kim2020, Kim2021} proposed limited feedback protocols within the ray-based channel model, utilizing only a subset of all the paths to optimize feedback utilization. Our prior work \cite{Lee2022} delved into the multipath Rician fading channel model, wherein each path is characterized by an independent Rician fading channel. We introduced several AoD-based zero-forcing (ZF) beamforming schemes tailored for environments with limited feedback.

While previous studies on CF-mMIMO with limited feedback often adopted equal feedback bit allocation among UEs or dominant paths \cite{Kim2020, Kim2021, Lee2022}, practical uplink feedback links are typically shared among UEs. Consequently, it's possible to more efficiently utilize the uplink feedback budget for each AP by adaptively allocating different feedback sizes among the UEs. Furthermore, the feedback bits assigned to each UE can be optimally distributed among the multiple paths within the channel. Hence, there's a necessity to explore the efficient allocation of feedback budget in FDD CF-mMIMO systems, particularly when channels encompass multiple resolvable paths with differing average powers.

In this paper, we propose user-centric UE-AP association and feedback bit allocation policies for the downlink of FDD CF-mMIMO systems with limited feedback, aimed at long-term efficacy. Given that the uplink control channel is typically shared among UEs, we consider each AP's total feedback budget, which is distributed among associated UEs. Employing the Saleh-Valenzuela channel model, which encompasses multi-resolvable paths with different average powers, we first identify the required feedback information for each UE, alongside developing an appropriate codebook structure. Subsequently, we formulate a joint optimization problem for user-centric UE-AP association and feedback bit allocation. Upon the allocation of feedback bits, each associated UE divides the assigned feedback size and utilizes a corresponding product codebook. Path gains are then quantized independently based on the feedback bit allocation outcome. We analyze the effects of feedback bit allocation on the sum rate and derive our proposed scheme from an alternative optimization problem, where the impacts of feedback bit allocation are taken into account. In the simulation part, we show that our proposed scheme effectively improves the performances of conventional approaches in CF-mMIMO systems.

\Blue
Our proposed scheme has two major contributions\KJcancel{ compared to existing works, which can be summarized as follows}:
\begin{itemize}
\item  Our proposed scheme distinguishes itself by incorporating the effects of imperfect CSI into the UE-AP association process—an aspect overlooked in the existing cell-free literature \cite{Buzzi2017, Buzzi2020, Bjornson2020, Kim2020, Kim2021}. Specifically, our approach adopts a user-centric UE-AP association strategy based on a unified metric that accounts for both large-scale fading coefficients and the performance degradation caused by imperfect CSI.

\item Unlike existing FDD cell-free literature \cite{Kim2020, Kim2021, Lee2022}, which either assumes equal average power across multiple resolvable paths at each UE or allocates feedback bits uniformly per UE, our proposed scheme optimally distributes feedback bits among UEs and their distinct resolvable paths, each characterized by different average power levels. This adaptive allocation maximizes the achievable sum rate by ensuring efficient utilization of the limited feedback resources in FDD CF-mMIMO systems.\KJcancel{ For example, the authors of \cite{Kim2020, Kim2021} assumed equal average power for all paths and allocated equal feedback bits to all UEs. Also, the authors of \cite{Lee2022} accounted for differing average powers across multiple paths by adopting a multipath Rician fading channel model but still employed equal feedback bit allocation. Meanwhile, our proposed scheme enables each AP to optimally allocate feedback bits to its associated UEs as well as their resolvable paths of different average powers. This} 
\end{itemize}

\Black
The rest of the paper is organized as follows: In Section \ref{problemFormulation}, we explain our system model and formulate our problem. In Section \ref{identification}, we identify the required feedback information for each UE, alongside developing an appropriate codebook structure. In Section \ref{quantizationEffect}, we analyze the effects of feedback bit splitting on the quantized channel and the achievable sum rate and establish alternative optimization problems. We propose our user-centric UE-AP association and feedback bit allocation in Section \ref{proposedScheme} and evaluate the performance of our proposed scheme in Section \ref{numericalResults}. Lastly, Section \ref{conclusion} concludes our paper.

\textbf{\emph{Notations.}}
For a vector $\x$, we denote by $\x^\dg$ its conjugate transpose and by $\Vert\x\Vert$ its $l_2$-norm. For an integer $n$, we use the notation $[n]$ to denote the set of all positive integers less than or equal to $n$, i.e., $[n]\teq\{1,\dots,n\}$. For given sets $\A$ and $\B$, we denote by $\vert\A\vert$ the cardinality of a set $\A$, and the expression $\A\setminus\B$ denotes the relative complement of $\B$ in $\A$, i.e., $\A\setminus\B\teq\{x\in\A:x\notin\B\}$.

\section{Problem Formulation\label{problemFormulation}}
In this section, we formulate our problem. We start by presenting an overview of our system model, followed by a detailed explanation of the setups. Subsequently, we delve into the description of our problem.

\begin{figure}[!t]
\centering
\includegraphics[width=0.8\columnwidth]{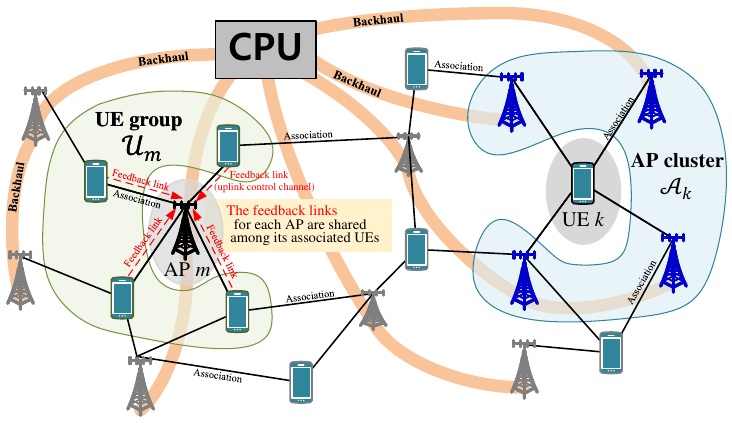}
\caption{\blue{An illustration of} our system model. \blue{Distributed APs are connected to the CPU via backhaul links, forming a CF-mMIMO system. The feedback links for each AP are shared among its associated UEs.}}
\label{fig1}
\vspace{-0.4cm}
\end{figure}

\subsection{System Model Overview}
Our system model is illustrated in Fig. \ref{fig1}. We consider a CF-mMIMO system operating with an FDD manner, where $M$ APs equipped with $N$ uniform linear array (ULA) antennas each and $K$ single-antenna UEs are randomly located on the network. All APs are connected to a central processing unit (CPU) through zero-delay and error-free backhauls, which enables data sharing among the APs. Thus, APs can serve multiple UEs in the same time-frequency band simultaneously. We adopt a user-centric approach for UE-AP association, where UEs initially determine the association of which detailed procedure will be explained in Section \ref{section:setup1}.

Once the UE-AP association is determined, APs serve their associated UEs accordingly. We represent the \emph{AP cluster} for UE $k$ as $\A_k\subset[M]$, where $k\in[K]$, comprising the APs associated with UE $k$. Similarly, $\U_m\subset[K]$ denotes the \emph{UE group} of AP $m$, where $m\in[M]$, comprising the UEs associated with AP $m$. For any UE-AP association, the AP clusters $(\A_1, \ldots, \A_K)$ determine the corresponding UE groups $(\U_1, \ldots, \U_M)$, and vice versa, whose relationship can be represented as follows:
\begin{align}
    \U_m&\triangleq \{k\in[K]: m\in\A_k\}, \quad \forall m\in[M],
    \label{eqn:UkAk}~\\
    \A_k&\triangleq \{m\in[M]: k\in\U_m\}, \quad \forall k\in[K].
    \label{eqn:AkUk}
\end{align}
Then, UE $k$'s received signal denoted by $y_k\in\C$ becomes
\begin{align}
  y_k
  &= \tsuml_{m\in\A_k} \h_{m,k}^\dg\x_m
     + \tsuml_{m\in\A_k^c} \h_{m,k}^\dg\x_m
     + n_k,\label{receivedSignal}
\end{align}
where $\h_{m,k}\in\C^{N\times1}$ represents the channel from AP $m$ to UE $k$, while $\x_m\in\C^{N\times1}$ denotes the transmitted signal from AP $m$. It satisfies the condition $\E[\mathrm{tr}(\x_m\x_m^\dg)]=P_m$, where $P_m$ denotes the transmit power budget of AP $m$. Also, $n_k\in\C$ is a circularly symmetric complex Gaussian noise at UE $k$ with zero mean and unit variance, i.e., $n_k\sim\mathcal{CN}(0, 1)$. In \magenta{the right-hand side (RHS) of} \eqref{receivedSignal}, the first term \KJcancel{$\sum_{m\in\A_k} \h_{m,k}^\dg\x_m$ }represents UE $k$'s received signal from its associated APs, encompassing both the desired signal and the same-group interference (SGI). The SGI arises from inter-UE interference within the same UE group. On the other hand, the second term \KJcancel{$\sum_{m\in\A_k^c} \h_{m,k}^\dg\x_m$ }corresponds to the received signal from APs outside the cluster, constituting inter-group interference (IGI).

As we stated earlier, we consider an FDD system, where the uplink and the downlink channels are not reciprocal. Thus, each UE should feed quantized channel information back to its associated APs to facilitate multiuser beamforming. In Section \ref{section:setup2}, we explain the details of our channel model and the limited feedback environment for each AP's CSI acquisition. After the feedback stage, each AP acquires the imperfect CSI of its associated UEs and determines suitable beamforming vectors to support them. We assume that each AP considers linear beamforming to support the associated UEs simultaneously; AP $m$'s \KJcancel{the }transmitted signal is constructed as \magenta{$\x_m=\sum_{j\in\U_m}\sqrt{p_{m,j}}\w_{m,j}s_j, $}\KJcancel{follows:
\begin{align}\label{transmittedSignal}
    \x_m=\textstyle\sum_{j\in\U_m}\sqrt{p_{m,j}}\w_{m,j}s_j,
\end{align}}
where $p_{m,j}$ is the transmit power allocated from AP $m$ to UE $j$, and $\w_{m,j}\in\C^{N\times1}$ is  AP $m$'s beamforming vector to serve UE $j$. Also, $s_j\in\C$ is the UE $j$'s data symbol such that $\E[\vert s_j\vert^2]=1$.

After a simple manipulation, \eqref{receivedSignal} can be rewritten as follows:
\begin{align}
    y_k
    &=\underbrace{\tsuml_{m\in\A_k} \sqrt{p_{m,k}} \h_{m,k}^\dg \w_{m,k}s_{k}}_{\textrm{desired~signal}}\NN
\end{align}
\begin{align}
    &\quad+\underbrace{\tsuml_{j\in[K]\setminus\{k\}}
    \tsuml_{m\in\A_j\cap\A_k}\sqrt{p_{m,j}} \h_{m,k}^\dg \w_{m,j} s_{j}}_\textrm{same-group~interference~(SGI)}\NNL
    &\quad+\underbrace{\tsuml_{j\in[K]\setminus\{k\}}
    \tsuml_{m\in\A_j\setminus\A_k}\sqrt{p_{m,j}}
    \h_{m,k}^\dg\w_{m,j}s_{j}}_\textrm{inter-group~interference~(IGI)}
    ~+~n_k. \label{rewrittenReceivedSignal}
\end{align}
In the \magenta{RHS} of \eqref{rewrittenReceivedSignal}, the first term corresponds to the desired signal of UE $k$, and the second term is the SGI, which is the interference caused by multiuser transmission within the same UE group. Also, the third term corresponds to the IGI, which is from the APs not belonging to the UE $k$'s AP cluster. We assume that each AP adopts the ZF beamforming to mitigate only the SGI among the associated UEs, allocating equal powers to them. Further details on the beamforming setup will be provided in Section \ref{section:setup3} along with the justification.

As a result of \eqref{rewrittenReceivedSignal}, the achievable rate of UE $k$ becomes
%
%
\begin{align}\label{achievableRate}
    &\R_k\teq\log_2
    \big(1+
        \D_k\big/
        (1+\I_k^\SGI+\I_k^\IGI)
    \big),
\end{align}
where $\D_k$, $\I_k^\SGI$, and $\I_k^\IGI$ denote the received signal powers of the desired signal, the SGI, and \KJcancel{the }IGI at UE $k$, respectively, given by
\begin{align}
    \D_k
    &\teq \big| \tsuml_{m\in\A_k}\sqrt{p_{m,k}} \h_{m,k}^\dg\w_{m,k}\big|^2, \label{DS}\NL
    \I_k^\SGI
    &\teq \tsuml_{j\in[K]\setminus \{k\}} \big| \tsuml_{m\in\A_j\cap\A_k} \sqrt{p_{m,j}} \h_{m,k}^\dg\w_{m,j} \big|^2, \label{SGI}\NL
    \I_k^\IGI
    &\teq\tsuml_{j\in[K]\setminus \{k\}} \big| \tsuml_{m\in\A_j\setminus\A_k} \sqrt{p_{m,j}} \h_{m,k}^\dg \w_{m,j} \big|^2 \label{IGI}.
\end{align}
Thus, the sum rate of the CF-mMIMO system becomes
\begin{align}\label{sumRate}
    \Rsum=\textstyle\sum_{k\in[K]}\R_k.
\end{align}

\subsection{Detailed Setups}\label{section:setup}
\subsubsection{User-Centric UE-AP Association}\label{section:setup1}
Many works on CF-mMIMO systems consider a scenario, where each UE is served by all APs in the network \cite{Ngo2017}. However, receiving data from APs in a far distance or bad conditions may not be efficient. Thus, in our system model, we consider that each UE is served by a subset of APs based on the UE-AP association result. Furthermore, we adopt a user-centric approach for UE-AP association, where each UE primarily determines the AP cluster (i.e., the set of APs) it belongs to.

We define $\Ao_k\subset[M]$ as the \emph{initial} AP cluster, a set of APs initially selected by UE $k$. Then, the initial AP clusters $(\Ao_1, \ldots, \Ao_K)$ determine the corresponding \emph{initial} UE groups $(\Uo_1, \ldots, \Uo_M)$ as in \eqref{eqn:UkAk}, where $\Uo_m\subset[K]$ is the initial UE group (i.e., the set of UEs) associated with AP $m$ given by \magenta{$\Uo_m=\{k\in[K]:m\in\Ao_k\}.$}
\KJcancel{\begin{align}
    \Uo_m=\{k\in[K]:m\in\Ao_k\}.
\end{align}
}However, the initial UE-AP association cannot always be valid due to the presence of unavailable APs in the initial AP clusters for several reasons. For example, the number of UEs simultaneously served by each AP can be limited by the number of transmit antennas. Even with a sufficient number of transmit antennas, the maximum number of serving UEs may still need to be constrained by backhaul capacity. Thus, it is reasonable to assume that the maximum number of serving UEs is regulated by a specific count at each AP. In this paper, we assume that the number of each AP's serving UEs cannot exceed $N_s$ such that $N_s\le N$, which means that each AP can transmit at most $N_s$ streams simultaneously.

Therefore, each AP (or the CPU) should determine (valid) UE-AP association $(\A_1, \ldots, \A_K)$ and $(\U_1, \ldots, \U_M)$ revising the initial association $(\Ao_1, \ldots, \Ao_K)$ and $(\Uo_1, \ldots, \Uo_M)$. To achieve this, each AP finds the UE groups $(\U_1, \ldots, \U_M)$, where for every $m\in[M]$, it is satisfied that $\U_m \subset \Uo_m $ and
\begin{align}
  \vert \U_m \vert = \min(\vert \Uo_m \vert,  N_s),
\end{align}
and then obtains the corresponding AP clusters $(\A_1, \ldots, \A_K)$ from \eqref{eqn:AkUk}. In the end, we introduce our user-centric association framework, which refines the initial user-centric UE-AP association by incorporating considerations for imperfect CSI. Throughout this paper, we operate under the assumption that the initial UE-AP association outcome is predetermined. If necessary (especially in the simulation part), we consider a simple initial UE-AP association scheme, wherein each UE selects a predetermined number of the nearest APs \cite{Buzzi2017, Buzzi2020}. However, it is important to note that our proposed framework is adaptable to any initial UE-AP association scheme.

\subsubsection{Channel Model and Limited Feedback Environment}\label{section:setup2}
For a channel model, we consider the Saleh-Valenzuela channel model, widely adopted for a multi-resolvable path environment \cite{Brady2013, Cheng2020, Kudathanthirige2024}. With the channel model, the vector channel from AP $m$ to UE $k$, i.e.,  $\h_{m,k}$ in \eqref{receivedSignal}, is represented by
\begin{align}\label{SV_Channel}
 \h_{m,k}
    &=\sqrt{\beta_{m,k}N}\Big[\textstyle\sum_{l\in[L]}
    \sqrt{\kappa\mkl}
    g\mkl\a(\theta\mkl)\Big],
\end{align}
where $\beta_{m,k}\in\mathbb{R}_+ \cup \{0\}$ is a large-scale fading coefficient between AP $m$ and UE $k$, and $L$ is the total number of dominant (line-of-sight or specular) paths. Also, $\kappa\mkI, \ldots, \kappa\mkL\in[0, 1]$ represent the relative strengths of a total $L$ channel components such that $\kappa\mkI + \dots + \kappa\mkL=1$, which we call \emph{dominance factors}. Furthermore, $g\mkl\sim\mathcal{CN}(0,1)$ and $\theta\mkl$ are the path gain and the AoD of the $l$th path from AP $m$ to UE $k$, respectively. Also, $\a(\theta)\in\C^{N\times 1}$ is the array steering vector corresponding to the AoD value $\theta$ given by \magenta{$\a(\theta) \triangleq \sqrt{(1/N)} [1,e^{j\frac{2\pi d}{\lambda}\mathrm{sin}\theta},\dots,e^{j(N-1)\frac{2\pi d}{\lambda}\mathrm{sin}\theta}]^T,$}
\KJcancel{\begin{align}\label{arraySteeringVector}
    &\a(\theta) \triangleq
    \sqrt{(1/N)}
    \big[1,e^{j\frac{2\pi d}{\lambda}\mathrm{sin}\theta},\dots,e^{j(N-1)\frac{2\pi d}{\lambda}\mathrm{sin}\theta}\big]^T,
\end{align}
}where $\lambda$ is the wavelength and $d$ is the antenna spacing, assumed to be half a wavelength, i.e., $d=\lambda/2$. As a result, it holds that $\E[\h\mk]=\mathbf{0}_N$ and $\E[\Vert\h\mk\Vert^2] = \beta\mk N$.

Note that in \eqref{SV_Channel}, the channel elements can be categorized into long-term varying and short-term varying ones based on their respective time scales; the long-term varying elements involve the large-scale fading coefficients, the dominance factors, and the AoDs
\magenta{(i.e., $\beta_{m,k}$, $\{\kappa\mkl\}_{l=1}^{L}$, and $\{\theta\mkl\}_{l=1}^L$)},
while the short-term varying elements include the path gains \magenta{(i.e., $\{g\mkl\}_{l=1}^L$)}.
\KJcancel{, i.e., in \eqref{SV_Channel}, \blue{long-term varying: $\beta_{m,k}, \{\kappa\mkl\}_{l=1}^{L}, \{\theta\mkl\}_{l=1}^L,$ short-term varying: $\{g\mkl\}_{l=1}^L.$}}
\KJcancel{\begin{align}
\textrm{\small Long-term varying:~~~~}
    &\beta_{m,k}, ~~\{\kappa\mkl\}_{l=1}^{L}, ~~\{\theta\mkl\}_{l=1}^L,\NNL
\textrm{\small Short-term varying:~~~~}
    &\{g\mkl\}_{l=1}^L. \NN
\end{align}
}\Blue
The time scales for the two categories of elements differ significantly: Short-term varying elements typically change on the order of a millisecond, while long-term varying elements change much more slowly—at least 40 times slower—even in high-mobility scenarios with UE speeds at 100 km/h and a carrier frequency of 2 GHz \cite{Ngo2017, Rappaport1956}.\KJcancel{ This results in a channel coherence time on the order of a millisecond, whereas long-term varying element changes are much less frequent. Consequently} \magenta{Therefore}, even in high-mobility environments, our protocol remains valid and can operate effectively with updates performed only a few times per second. This approach aligns with other CF-mMIMO studies that employ long-term strategies for resource allocation \cite{Ngo2017, Buzzi2017, Buzzi2020, Bjornson2020}.

\Black
We proceed with the assumption that \blue{all the long-term varying elements are known a priori, wherever required (e.g., at the CPU, APs, or UEs), since these elements are easy to measure and can be shared among APs \cite{Lee2016, Sheng2011, Hoon2014, Ngo2017, Buzzi2017, Buzzi2020, Bjornson2020}. Each AP and each UE can individually estimate these long-term varying elements by estimating the channel covariance matrix \cite{Kim2020, Abdallah2020} or calculating the Friis transmission equation \cite{Friis1946}. We further assume that the large-scale fading coefficients, dominance factors, and AoA/AoD values estimated by the AP and UE are identical.} This assumption is based on the fact that these long-term varying elements are reciprocal between the uplink and downlink channels even in the FDD systems \blue{\cite{Kim2020, Kim2021, Abdallah2020, Lee2022}. Specifically, in the AP perspective,} AP $m$ directly measures the long-term varying elements from the \emph{uplink} channel of UE $k$ (i.e., $\beta_{m,k}$, $\{\kappa\mkl\}_{l=1}^{L}$, and $\{\theta\mkl\}_{l=1}^L$), and performs the same measurement over all UEs' uplink channels. Then, APs share these values via the connected CPU so that each AP becomes aware of
\begin{align}\label{LTV_elements}
  \big\{\beta_{m,k}, ~~\kappa\mkl, ~~\theta\mkl
  :~ m\in[M], k\in[K], l\in[L]
  \big\}.
\end{align}
\blue{This process is similarly performed by each UE, enabling every UE to individually acquire all the long-term varying elements listed in \eqref{LTV_elements}.}

To help the beamforming design, each UE feeds the quantized information about the short-term varying elements to every associated AP. Considering the fact that the uplink feedback links are typically shared among the associated UEs for each AP \cite{Yoo2006, Lee2013}, we assume that AP $m$ has a total of $B_m$-bit feedback link that is exclusively shared among its associated UEs (i.e., $\U_m$). Thus, we have the feedback link sharing constraints for AP $m$ as follows:
\begin{align}
    &\textstyle\sum_{k\in\U_m} b_{m,k}=B_m,\\
    &~b_{m,k}\ge 0, \quad \text{for all}~~k\in\U_m,\\
    &~b_{m,k}=0,  \quad \text{for all}~~k\not\in\U_m,
\end{align}
where $b_{m,k}$ is allocated feedback bits from AP $m$ to UE $k$. \blue{We will describe the codebook structure considered in our paper in Section \ref{codebookStructure}.}

\subsubsection{ZF Beamforming and Achievable Rate}\label{section:setup3}
From a beamforming perspective, each AP requires only the CSI of its associated UEs for SGI mitigation, which is easily obtained through the desired channels feedback from those UEs. However, for IGI mitigation, each AP also needs to know the interfering channels of unassociated UEs, which is considerably challenging. Thus, we assume each AP's beamforming focuses on mitigating SGI using only the CSI of associated UEs.

To nullify SGI, each AP adopts the ZF beamforming; with the perfect CSI of the associated UEs, AP $m$'s ZF beamforming vector for UE $k\in\U_m$  denoted by $\w_{m,k}^\star$ becomes a unit vector chosen from the null space of the other associated UEs' channels, so it is satisfied that $\Vert\w_{m,k}^\star\Vert^2=1$ and \magenta{$\w_{m,k}^\star\perp\{\h_{m,j}:j\in\U_m\setminus\{k\}\}.$}
\KJcancel{\begin{align}\label{ideal_ZF_beamforming}
    \w_{m,k}^\star\perp\big\{\h_{m,j}:j\in\U_m\setminus\{k\}\big\}.
\end{align}
}Then, the SGI term in \eqref{rewrittenReceivedSignal} is nullified, i.e., $\I_k^\SGI=0$. With the imperfect CSI, however, the ZF beamforming vector becomes a unit vector such that
\begin{align}\label{practical_ZF_beamforming}
    \w_{m,k}\perp\{\hh_{m,j}:j\in\U_m\setminus\{k\}\},
\end{align}
where $\hh_{m,j}\in\C^{N\times 1}$ the imperfect version of $\h_{m,j}$ possessed at AP $m$. Thus, the SGI term in \eqref{rewrittenReceivedSignal} still remains due to the quantization error, i.e., $\I_k^\SGI\ne0$.

Meanwhile, to focus on UE-AP association and feedback bit allocation, we assume that each AP serves the associated UEs with equal powers, i.e., $p_{m,k}=p_m$ for all $k\in\U_m$, where
\begin{align}
    p_m\triangleq {P_m}/{\vert\U_m\vert}.
    \label{eqn:EP}
\end{align}
Note that UE-AP association and feedback bit allocation are relatively long term policies compared to power allocation based on instantaneous channel gain feedback. Thus, in this paper, we tackle UE-AP association and feedback bit allocation separately from power allocation. Although we consider equal power allocation, our framework can accommodate other power allocation schemes. Moreover, equal power allocation is frequently considered in FDD CF-mMIMO systems (e.g., over multiple beams \cite{Kim2020} or multiple UEs \cite{Zhang2024}) to concentrate the efficient use of limited feedback. Meanwhile, ZF beamforming with equal power allocation is known as optimal in the high signal-to-noise ratio (SNR) region in MIMO broadcast channels \cite{Yoo2006, Lee2015}, and the performance in the low SNR region can be enhanced by employing regularized ZF beamforming (or minimum mean square error beamforming) \cite{Jindal2006, Lee2013}.

\subsection{Problem Description}
Our objective is to determine long-term policies for achieving the optimal user-centric UE-AP association and feedback bit allocation, aiming to maximize the sum rate, which can be formulated as follows:
\begin{align}
  &\PAP{1}
  \hspace{-.1in}
  \underset{\{\U_m \subset\Uo_m\}_{m\in[M]}
  \atop  \{b_{m,k}\}_{m\in[M], k\in[K]}}
  {\mathrm{maximize}}
    \Rsum \NNL
    &\mathrm{subject~to}~~
     \vert\U_m\vert = \min(\Uo_m, N_s),
     \quad
     \forall m\in[M] \label{constraint0}\\
    &\qquad\qquad~~
        \textstyle\sum_{k\in \U_m} b_{m,k}=B_m,
        \quad~
        \forall m\in[M] \label{constraint1}\\
    &\qquad\qquad~~~
        b_{m,k}\ge 0,
        \qquad
        \forall k\in\U_m~~\forall m\in[M] \label{constraint2}\\
    &\qquad\qquad~~~
        b_{m,k}=0,
        \qquad
        \forall k\not\in\U_m~~\forall m\in[M]. \label{constraint3}
\end{align}
Note that the problem \PAP{1} is hard to solve directly due to the implicit incorporation of feedback bit allocation effects within the objective function. Additionally, the association problem entails a mixed-integer nature, a characteristic commonly acknowledge to be NP-hard.

\Blue
\KJcancel{Once the UE-AP association is determined, each AP finds ZF beamforming vectors to mitigate SGI. }From the UE-AP association perspective, maximizing the achievable sum rate requires determining associations that not only enhance the desired signal power but also account for both manageable SGI and unmanageable IGI. From the feedback bit allocation perspective, the AP should allocate feedback bits to improve the channel accuracy for SGI, thereby further increasing the achievable sum rate. In \KJcancel{practical implications, this}\magenta{this context, the} problem \magenta{\PAP{1}} implies that UE-AP association and feedback bit allocation should be optimized to ensure efficient feedback usage, balancing the tradeoffs between the number of associated UEs and channel accuracy necessary to manage SGI.

\Black
In the next sections, we delineate the feedback information pertinent to each UE alongside the suitable codebook structure. Furthermore, we conduct an analysis on the ramifications of feedback bit allocation for any valid UE-AP association.

\section{Identification of Feedback Information and Codebook Structure\label{identification}}
In this section, we elaborate on the optimal utilization of allocated feedback bits by each UE. We delineate the pertinent feedback information for individual UEs and establish the structure of the codebook accordingly.

\subsection{Identification of Feedback Information}
\blue{We recall the channel from AP $m$ to UE $k$ such that $k\in\U_m$ given in \eqref{SV_Channel}.} As mentioned in Section \ref{section:setup2}, AP $m$ can directly measure the long-term varying elements in the channel \emph{from} the UE $k$, which are $\beta_{m,k}$, $\{\kappa\mkl\}_{l=1}^{L}$, and $\{\theta\mkl\}_{l=1}^L$ in \eqref{SV_Channel}. Thus, to assist the AP $m$'s reconstruction of $\h_{m,k}$, UE $k$'s feedback may include the information about the short-term varying elements: $\{g\mkl\}_{l=1}^L$. In our case, \KJcancel{UE $k$  provides feedback only on the quantized information of $\{g\mkl\}_{l=1}^L$ to AP $m$.

In summary, }each UE provides feedback information to every associated AP on the $L$ path gains; UE $k$ feeds back the information on $\{g\mkI, \ldots, g\mkL\}$ to AP $m$ with a total of $b_{m,k}$-bit feedback for every $m\in\A_k$.

\subsection{Codebook Structure\label{codebookStructure}}
Now, we explain each UE's codebook structure to quantize $L$ dominant path gains with the allocated feedback bits.

In much of the literature concerning limited feedback in MIMO systems, each UE typically quantizes solely the channel direction, represented as a unit vector, to aid in AP's beamforming \cite{Jindal2006, Lee2016}. For multipath channel models, limited feedback protocols have been proposed in \cite{Kim2020, Kim2021, Lee2022}, where each UE feeds back the direction of a vector comprising the gains of multiple paths, not fully considering the different average gains among the paths. However, in our scenario, each UE directly quantifies the path gains with different dominance factors rather than considering the direction, rendering the aforementioned limited feedback approach unsuitable.

In each channel, the gains of $L$ dominant paths are i.i.d. circularly symmetric complex Gaussian random variables with $\CN(0,1)$, which means that each channel has a total of $2L$ i.i.d. real-valued Gaussian random variables with $\mathcal{N}(0, 1/2)$ to quantize. Thus, we consider a product codebook structure comprising $2L$ scalar codebooks, where the $2L$ elements are quantized independently. While vector quantization generally outperforms scalar one, the latter is simpler and offers greater flexibility for adjustments, making it appropriate in our case.

When UE $k$ quantizes $\{g\mkl\}_{l=1}^L$ with a total of $b_{m,k}$ bits, UE $k$ splits $b\mk$ into $(b\mkI, \ldots, b\mkL)$ such that $\sum_{l=1}^L b\mkl = b_{m,k}$ and quantizes each of $\{g\mkl\}_{l=1}^L$ with $b\mkl$ bits. In this case, the $l$th path gain can be divided into real and imaginary parts as \magenta{$g\mkl=\Re{g\mkl}+j\Im{g\mkl}.$}
\KJcancel{\begin{align}\label{dividedPathGain}
    g\mkl=\Re{g\mkl}+j\Im{g\mkl}.
\end{align}
}Since they are i.i.d. random variables with $\mathcal{N}(0, 1/2)$ and suffer from the same dominance factor (e.g., $\kappa\mkl$), we assume that each part is quantized with the half of $b\mkl$ bits.

\subsection{Each UE's Feedback Bit Splitting and Reconstructed Channels from Feedback Information}
Once the feedback sizes of  UE $k$ (i.e., $\{b\mk\}_{m\in\A_k}$) are determined from the solution of the problem \PAP{1}, UE $k$ solves the following feedback bit splitting problem:
\begin{align}
  \PUE{1}
  \underset{\{b\mkl\}_{m\in\A_k, l\in[L]}}
  {\mathrm{maximize}}
    \R_k \hspace{-.5in}&\NNL
    \mathrm{subject~to}~~
    &\textstyle\sum_{l\in[L]} b\mkl =b_{m,k}, ~~\forall m\in\A_k\label{constraint_k1}\\
    &b\mkl\ge 0,  \qquad\forall m\in\A_k ~\forall l\in[L].
    \label{constraint_k2}
\end{align}
Then, for every $m\in\A_k$, UE $k$ quantizes $\{g\mkl\}_{l=1}^L$ into $\{\hg\mkl\}_{l=1}^L$, where $\hg\mkl$ is the $b\mkl$-bit quantized version of $g\mkl$. Thus, after the feedback stage, AP $m$ can construct the quantized version of $\h\mk$ from feedback information as follows (cf., \eqref{SV_Channel}).
\begin{align}\label{quantizedChannel}
 \hh_{m,k}
    &=\sqrt{\beta_{m,k}N}
    \Big[\textstyle\sum_{l\in[L]}
    \sqrt{\kappa\mkl}
    \hg\mkl\a(\theta\mkl)\Big].
\end{align}

\begin{remark}
One may wonder how each AP can know the feedback splitting result at each UE. It is worth noting that through a reframing of the original problem \PAP{1}, the allocation of feedback bits to UEs will be detailed in Sections \ref{PAP3b_solution} and \ref{PUE2_solution}. This allocation will primarily rely on the long-term varying elements specified in \eqref{LTV_elements}. Furthermore, each UE's feedback bit splitting will be designed only with its own long-term varying elements. Therefore, once the long-term varying elements are given, each AP (or the CPU) can directly calculate every AP's feedback bit allocation to its associated UEs as well as every UE's feedback bit splitting. Therefore, each AP only needs to convey the allocated feedback size to its associated UE, which is a single scalar value. With the allocated feedback size (e.g., $b\mk$), both the AP and the UE know how it is split for $L$ dominant paths (e.g., $\{b\mkl\}_{l=1}^L$). Further details will be provided in the following sections.
\end{remark}

\section{Effects of Channel Quantization and Alternative Optimization Problems\label{quantizationEffect}}
In this section, we analyze the effects of quantization errors on the achievable rate for  an arbitrary \emph{valid} UE-AP association. Based on the analysis, we establish alternative problems to figure out the optimal user-centric association and feedback bit allocation for long-term policies.

\subsection{The Effects of Feedback Bit Splitting on the Quantized Channel via Rate-Distortion Quantizer}
We first explain how the feedback bit splitting in each UE affects the quantized channel via a rate-distortion quantizer. In information theory, rate-distortion theory establishes the relationship between the encoding size and the achievable distortion in lossy data compression \cite{Cover1999}. We begin with Lemma 1, a well-known result from rate-distortion theory.
\begin{lemma}(The rate-distortion for Gaussian sources \cite{Cover1999})
When compressing a memoryless source of a Gaussian random variable $X$ with variance $\sigma_X^2$ into $\hat{X}$, the minimum (compressing) rate $R$ to achieve the distortion $D \teq \E[(X-\hat{X})^2]$ is
\begin{align}\label{RDF}
R(D)=
\begin{cases}
    (1/2)\log_2(\sigma_X^2/D),&0\leq D\leq\sigma_X^2~\\
    0,&D>\sigma_X^2.
\end{cases}
\end{align}
\end{lemma}

Now, we consider the feedback bit splitting at UE $k$ when quantizing the channel from AP $m$ with $b\mk$ bits. We revisit the channel between AP $m$ and UE $k$ (i.e.,  $\h_{m,k}$ given in \eqref{SV_Channel}) and its quantized version (i.e., $\hh_{m,k}$ given in \eqref{quantizedChannel}). Then, the channel quantization error denoted by $\e_{m,k}\in\C^{N\times 1}$ becomes
\begin{align}
    \e_{m,k}&\teq\h_{m,k}-\hh_{m,k}~\NNL
    &=\sqrt{\beta_{m,k}N}\Big[\textstyle\sum_{l\in[L]}\sqrt{\kappa\mkl}
    \Delta\mkl\mathbf{a}(\theta\mkl)\Big],
    \label{resultingErrorVector}
\end{align}
where $\Delta\mkl(\teq g\mkl-\hg\mkl)$ is the path gain quantization error at the $l$th path when quantizing $g\mkl$ with $b\mkl$ bits.

For analytic tractability, we assume that each UE uses the rate-distortion quantizer to quantize each path gain, which can achieve the rate-distortion function given in \eqref{RDF}. Then, we obtain the following lemma.
\begin{lemma}
With the rate-distortion quantizer, the path gain quantization error at the $l$th path (i.e., $\Delta\mkl$ in \eqref{resultingErrorVector}) follows a circularly symmetric complex Gaussian distribution with zero mean and variance $2^{-b\mkl}$, i.e., $\Delta\mkl \sim \mathcal{CN}(0, 2^{-b\mkl})$.
\end{lemma}
\begin{IEEEproof}
In our case, $g\mkl$ is a circularly symmetric Gaussian random variable of $\mathcal{CN}(0, 1)$, so each of its real and imaginary parts is $\mathcal{N}(0, 1/2)$. When quantizing $g\mkl$ with $b\mkl$ bits, each part is quantized with $b\mkl/2$ bits, which corresponds to the case with $\sigma_X^2=1/2$ and $R(D)=b\mkl/2$ in Lemma 1. Thus, the quantization error at each part (i.e., $\Re{\delta\mkl}$ or $\Im{\delta\mkl}$) becomes an i.i.d. random variable with $\mathcal{N}(0, 2^{-b\mkl-1})$, and hence the quantization error $\Delta\mkl$ follows circularly symmetric complex Gaussian distribution of $\mathcal{CN}(0, 2^{-b\mkl})$.
\end{IEEEproof}

Thus, with the feedback bit splitting $(b\mk^{\scriptscriptstyle(1)}, \ldots, b\mk^{\scriptscriptstyle(L)})$ such that $\sum_{l\in[L]}b\mkl=b\mk$, the quantized channel $\hh\mk$ becomes
\begin{align}
    \hh_{m,k}
    &\!=\!\!\sqrt{\beta_{m,k}}
    \Big[\textstyle\sum_{l\in[L]}
    \sqrt{\kappa\mkl}
    (g\mkl-\Delta\mkl)\a(\theta\mkl)\Big],
\end{align}
where $\Delta\mkl\sim\mathcal{CN}(0, 2^{-b\mkl})$.

\subsection{The Effects of Quantization Errors on the Achievable Rate}
We then explain how the quantization errors affect the achievable rate. We revisit the achievable rate of UE $k$ given in \eqref{achievableRate} along with \eqref{DS}--\eqref{IGI} and the equal power allocation \eqref{eqn:EP}. In the following lemma, we find the upper bounds of $\E_{\h,\mathcal{C}}[\D_k]$, $\E_{\h,\mathcal{C}}[\I_k^\SGI]$, and $\E_{\h,\mathcal{C}}[\I_k^\IGI]$, which are the desired signal, the SGI, and the IGI powers averaged on short-term varying channels and quantization codebooks, respectively.
\begin{lemma}
The terms $\E_{\h,\mathcal{C}}[\D_k]$, $\E_{\h,\mathcal{C}}[\I_k^\SGI]$, and $\E_{\h,\mathcal{C}}[\I_k^\IGI]$ are upper bounded as $\E_{\h,\CB}[\D_k] \le\D_k^{\downarrow}$, $\E_{\h,\CB} \big[\I_k^\SGI\big] \le\I_k^{\SGI\downarrow}$, and $\E_{\h,\CB} \big[\I_k^\IGI\big] \le\I_k^{\IGI\downarrow}$, where
\begin{align}
&\D_k^{\downarrow}\teq
    \tsuml_{m\in\A_k}\pmk\beta_{m,k}N
\label{eqn:D_k}\\
&\I_k^{\SGI\downarrow} \triangleq
    \tsuml_{\substack{j\in[K]    \\   j\ne k}}
    \tsuml_{\substack{m\in\A_j   \\   m\in\A_k}}
    \pmk\beta_{m,k}N
    \Big[
    \sum\limits_{l\in[L]}
        \kappa\mkl 2^{-b\mkl} \Big]
    \label{eqn:I_k_SGI}\\
    &\I_k^{\IGI\downarrow}\teq
    \tsuml_{\substack{j\in[K]    \\   j\ne k}}
    \tsuml_{\substack{m\in\A_j   \\   m\notin\A_k}}
        \pmk\beta_{m,k}N.
    \label{eqn:I_k_IGI}
\end{align}
\end{lemma}
\begin{IEEEproof}
See Appendix A.
\end{IEEEproof}

In the objective functions of the problems \PAP{1} and \PUE{1}, the effects of feedback bit allocation/splitting are implicit. To further explore long-term strategies for UE-AP association and feedback bit allocation, we leverage Lemma 3 to formulate an alternative achievable rate $\R_k'$ as follows:
%
\begin{align}\label{eqn:Rk'}
    \R_k'\teq
    \log_2\big(1+
    \D_k^{\downarrow}\big/
    (1+\I_k^{\SGI\downarrow}+\I_k^{\IGI\downarrow})\big),
\end{align}
where the effects of feedback bit allocation/splitting are explicit, averaged over all possible quantization realizations. This comprehensive approach is essential for a long-term policy, which must consider varying elements over time and reflect the averaged effects of short-term fadings and quantizations.

\Blue
Meanwhile, the alternative sum rate derived from \eqref{eqn:Rk'} is $\sum_{k\in[K]}\R_k'$, but the impact of feedback bit allocation on $\sum_{k\in[K]}\R_k'$ is difficult to analyze directly.
Thus, for analytic tractability, we find a lower bound on $\sum_{k\in[K]}\R_k'$ as follows:
\begin{align}
\tsuml_{k\in[K]}\R_k'
    &=\tsuml_{k\in[K]}
    \log_2\Big(1+\D_k^{\downarrow}+\I_k^{\SGI\downarrow}+\I_k^{\IGI\downarrow}\Big)\NNL
    &~~~-\tsuml_{k\in[K]}
    \log_2\Big(1+\I_k^{\SGI\downarrow}+\I_k^{\IGI\downarrow}\Big)\NNL
    &\stackrel{(a)}{\ge}
    \tsuml_{k\in[K]}
    \log_2\Big(1+\D_k^{\downarrow}
    +\I_{k, \lbound}^{\SGI\downarrow}
    +\I_k^{\IGI\downarrow}\Big)\NNL
    &~~~-K\cdot\log_2\Big(1+\tfrac{1}{K}\tsuml_{k\in[K]}
    \big(\I_k^{\SGI\downarrow}+\I_k^{\IGI\downarrow}\big)\Big),\label{eqn:R'_sum}
\end{align}
where $\I_{k, \lbound}^{\SGI\downarrow}$ is the lower bound of $\I_k^{\SGI\downarrow}$ given by
%
\begin{align}
\I_{k,\lbound}^{\SGI\downarrow} \triangleq
    \tsuml_{\substack{j\in[K]    \\   j\ne k}}
    \tsuml_{\substack{m\in\A_j   \\   m\in\A_k}}
    \pmk\beta_{m,k}N
    \Big[
    \tsuml_{l\in[L]}
        \kappa\mkl 2^{-B_m} \Big],\NN
\end{align}
%
which is obtained by assuming that the entire feedback budget of AP $m$ (i.e., $B_m$) is allocated to every channel path (i.e., $b\mkl$) in \eqref{eqn:I_k_SGI}.
%
%
In \eqref{eqn:R'_sum}, the inequality $(a)$ holds due to the relationship $\I_{k} ^{\SGI\downarrow} \ge \I_{k, \lbound}^{\SGI\downarrow}$ and Jensen's inequality.

Thus, we consider the alternative sum rate $\Rsum'$ such that $\Rsum' \le \sum_{k \in [K]} \R_k'$ defined by
\begin{align}
    \Rsum'&\teq\tsuml_{k\in[K]}
    \log_2\Big(1+\D_k^{\downarrow}
    +\I_{k,\lbound}^{\SGI\downarrow}
    +\I_k^{\IGI\downarrow}\Big)\NNL
    &~~~-K\cdot\log_2\Big(1+\tfrac{1}{K}\tsuml_{k\in[K]}
    \big(\I_k^{\SGI\downarrow}+\I_k^{\IGI\downarrow}\big)\Big).
    \label{eqn:Rsum'}
\end{align}
\KJcancel{As can be seen in \eqref{eqn:D_k}--\eqref{eqn:I_k_IGI}, the feedback bit allocation is solely related to $\I_k^{\SGI\downarrow}$ in \eqref{eqn:Rsum'}. Consequently, in \eqref{eqn:Rsum'}, the feedback bit allocation impacts only the term $\sum_{k\in[K]} \I_k^{\SGI\downarrow}$, representing the overall SGI caused by quantization errors.
Although in \eqref{eqn:Rsum'} we consider the averaged SGI across all serving UEs rather than the individual SGI of each UE, this does not imply that each UE experiences, or is expected to experience, a similar level of SGI. Instead, it indicates that we focus on utilizing the limited feedback budget to minimize the overall interference among the associated UEs caused by quantization errors.}
\KJcancel{In the simulation part, we compare the alternative sum rate \eqref{eqn:Rsum'} with the actual sum rate, demonstrating that the alternative sum rate closely approximates the actual one.
The expanded versions of \eqref{eqn:Rk'} and \eqref{eqn:Rsum'} are respectively given in \eqref{eqn:Rk'_full} and \eqref{eqn:Rsum'_full}.}

\KJcancel{
\begin{figure*}[t]\small\Red
\begin{align}
\R_k'&\teq
\log_2\Bigg(1+
    \frac{\sum\limits_{m\in\A_k}\pmk\beta_{m,k}N}
    {1+
    \sum\limits_{\substack{j\in[K]    \\   j\ne k}}
        \sum\limits_{\substack{m\in\A_j   \\   m\in\A_k}}
        \pmk\beta_{m,k}N
        \Big[
            \sum\limits_{l\in[L]}
            \kappa\mkl 2^{-b\mkl}
        \Big]
    +
    \sum\limits_{\substack{j\in[K]    \\   j\ne k}}
    \sum\limits_{\substack{m\in\A_j   \\   m\notin\A_k}}
    \pmk\beta_{m,k}N}\Bigg),\label{eqn:Rk'_full}\\
\Rsum'&\teq
\sum\limits_{k\in[K]}
\log_2\Bigg(1+\sum\limits_{m\in\A_k}\pmk\beta_{m,k}N
\blue{\KJcancelc{+\sum\limits_{\substack{j\in[K]    \\   j\ne k}}
        \sum\limits_{\substack{m\in\A_j   \\   m\in\A_k}}
            \pmk\beta_{m,k}N
            \Big[
                \sum\limits_{l\in[L]}
                \kappa\mkl 2^{-B_m}
            \Big]}}
+\sum\limits_{\substack{j\in[K]    \\   j\ne k}}
      \sum\limits_{\substack{m\in\A_j   \\   m\notin\A_k}}
    \pmk\beta_{m,k}N\Bigg)\NNL
&~~~-K\cdot\log_2\Bigg(1+\frac{1}{K}\sum_{k\in[K]}\Bigg(
    \sum\limits_{\substack{j\in[K]    \\   j\ne k}}
        \sum\limits_{\substack{m\in\A_j   \\   m\in\A_k}}
            \pmk\beta_{m,k}N
            \Big[
                \sum\limits_{l\in[L]}
                \kappa\mkl 2^{-b\mkl}
            \Big]
    + \sum\limits_{\substack{j\in[K]    \\   j\ne k}}
      \sum\limits_{\substack{m\in\A_j   \\   m\notin\A_k}}
    \pmk\beta_{m,k}N\Bigg)\Bigg).\label{eqn:Rsum'_full}
\end{align}
\hrule
\end{figure*}
}

\Black

\subsection{Alternative Optimization Problems}
Now, we are ready to establish alternative optimization problems based on $\R_k'$ in \eqref{eqn:Rk'} and $\Rsum'$ in \eqref{eqn:Rsum'}. Each AP tries to maximize $\Rsum'$ instead of $\Rsum$ in the problem \PAP{1}, while each UE tries to maximize $\R_k'$ instead of $\R_k$ in the problem \PUE{1}. In other words, each AP (or the CPU) solves the following alternative optimization problem:
\begin{align}
  \PAP{2}~~~
  \underset{\{\U_m \subset\Uo_m\}_{m\in[M]}
  \atop  \{b_{m,k}\}_{m\in[M], k\in[K]}}
  {\mathrm{maximize}}
    ~~&\Rsum' \NNL
    \mathrm{subject~to}~~~~~
    &\eqref{constraint0}, \eqref{constraint1}, \eqref{constraint2}, \eqref{constraint3}.\NN
\end{align}
Also, UE $k$ solves the following alternative problem:
\begin{align}
  \PUE{2}~
  \underset{\{b\mkl\}_{m\in\A_k, l\in[L]}}
  {\mathrm{maximize}}
  &
  ~~\R_k'
 \NNL
    \mathrm{subject~to}~~
    &\eqref{constraint_k1}, \eqref{constraint_k2}.~~~\NN
\end{align}
Note that when each AP solves \PAP{2}, the feedback bit splitting at each UE should be considered as it affects the objective function (see \eqref{eqn:Rsum'}\KJcancel{\eqref{eqn:Rsum'_full}}). Thus, each AP should solve \PAP{2} along with \PUE{2} for all $k\in[K]$.

Meanwhile, we obtain the following observation.
\begin{observation}
In the problems \PAP{2} and \PUE{2}, we can observe the following facts:
\begin{itemize}
\item In \eqref{eqn:D_k}--\eqref{eqn:I_k_IGI},  only the term $\I_k^{\SGI\downarrow}$ is dependent on $b\mkl$, while $D_k^{\downarrow}$ and $\I_k^{\IGI\downarrow}$ are not.

\item The term $\I_k^\SGIDA$ given in \eqref{eqn:I_k_SGI} is a convex function of $b\mkl$ for any $(m,k,l)$.

\item From the feedback bit allocation/splitting perspective, maximizing $\R_k'$
     (i.e., \eqref{eqn:Rk'}\KJcancel{ and \eqref{eqn:Rk'_full}}) is equivalent to minimizing $\I_k^\SGIDA$, while maximizing $\Rsum'$ (i.e., \eqref{eqn:Rsum'}\KJcancel{ and \eqref{eqn:Rsum'_full}}) is equivalent to minimizing $\sum_{k\in[K]} \I_k^\SGIDA$.

\item The constraint functions \eqref{constraint1} and \eqref{constraint_k1} have the relationship such that
    \begin{align}
     \tsuml_{k\in \U_m} b_{m,k}
     = \tsuml_{k\in \U_m} \Big[\tsuml_{l\in[L]} b\mkl\Big]
     = B_m,
     \quad \forall m\in[M].\NN
    \end{align}
\end{itemize}
\end{observation}

\magenta{From Observation 1, although in \eqref{eqn:Rsum'} we consider the averaged SGI across all serving UEs rather than the individual SGI of each UE, this does not imply that each UE experiences, or is expected to experience, a similar level of SGI. Instead, it indicates that we focus on utilizing the limited feedback budget to minimize the overall interference among the associated UEs caused by quantization errors.}

Observation 1 renders it adequate for each AP to solve a single composite problem:
\begin{align}
  &\PAP{3}~
  \underset{\{\U_m \subset\Uo_m\}_{m\in[M]}
  \atop  \{b\mkl\}_{m\in[M], k\in[K], l\in[L]}}
  {\mathrm{maximize}}
    ~~\Rsum' \NNL
    &~~~\mathrm{s.t.}~~
     \vert\U_m\vert = \min(\Uo_m, N_s),
     \quad
     \forall m\in[M] \label{cconstraint0}\\
    &\qquad\quad
        \tsuml_{k\in \U_m} \tsuml_{l\in[L]}b\mkl=B_m,
        \quad
        \forall m\in[M] \label{cconstraint1}\\
    &\qquad\quad~
        b\mkl\ge 0,
        \quad
        \forall m\in[M]~~\forall k\in\U_m ~~\forall l\in[L] \label{cconstraint2}~\\
    &\qquad\quad~
        b\mkl=0,
        \quad
        \forall m\in[M]~~\forall k\not\in\U_m ~~\forall l\in[L] . \label{cconstraint3}
\end{align}
Once the solution of \PAP{3} is obtained as $\{\U_m^\star\}_{m\in[M]}$ and $\{b\mkls\}_{m\in[M], k\in[K], l\in[L]}$, the solution of the \PAP{2} becomes $\{\U_m^\star\}_{m\in[M]}$ and $\{b\mk^\star\}_{m\in[M], k\in[K]}$, where $b\mk^\star\triangleq \sum_{l\in[L]}b\mkls$. Also, the solution of the problem \PUE{2} becomes $\{b\mkls\}_{m\in\A_k^\star, l\in[L]}$, where $\A_k^\star$ is the AP cluster for UE $k$ corresponding to the UE grouping result $\{\U_m^\star\}_{m\in[M]}$.

\begin{remark}
Once the feedback bit allocation to UEs (i.e., $b\mk$) is announced, UE $k$ can solve \PUE{2} using only its \emph{own} long-term varying elements as shown in \eqref{eqn:Rk'}\KJcancel{\eqref{eqn:Rk'_full}}. Note that each AP can do the same task, being aware of what UE $k$ knows.
\end{remark}

\section{Our Proposed User-Centric Association and Feedback Bit Allocation Protocol\label{proposedScheme}}
In this section, we explain our proposed user-centric UE-AP association and feedback bit allocation protocol. We first explain the procedure of our proposed protocol and provide the solutions of the problems \PAP{3} and \PUE{2}.

\subsection{Procedure of Our Proposed Protocol\label{procedure}}
Our proposed association and feedback bit allocation protocol is based on the procedure to solve the problem \PAP{3}, which can be summarized as follows.
\begin{itemize}
    \item Stage 0: At every AP, the initial user-centric UE-AP association is primarily given, i.e., $(\Uo_1, \ldots, \Uo_M)$, and the corresponding $(\Ao_1, \ldots, \Ao_K)$ are given.

    \item Stage 1: Each AP solves \PAP{3} and obtains the UE-AP association with the feedback bit allocation/splitting results. Consequently, each AP knows $(\U_1, \ldots, \U_M)$ alongside their respective $(\A_1, \ldots, \A_K)$ and $\{b\mkl\}_{m\in[M], k\in[K], l\in[L]}$ as a solution of \PAP{3}.

    \item Stage 2: \blue{Each AP broadcasts the association result along with a \emph{single} allocated feedback size for each associated UE (e.g., $b\mk$).} Thus, each UE becomes aware of its own AP cluster as well as the allocated feedback bits. For example, AP $m$ informs every UE $k$ within its set $\U_m$ of the association result as well as the allocated feedback bits $b\mk(\triangleq \sum_{l\in[L]}b\mkl)$.

    \item Stage 3: Each UE solves \PUE{2} and splits the allocated feedback bits. Then, each UE quantizes path gains with the corresponding split feedback bits. For example, UE $k$ finds $\{b\mkl\}_{m\in\A_k, l\in[L]}$ from $\{b\mk\}_{m\in\A_k}$ and quantizes $\{g\mkl\}_{m\in\A_k, l\in[L]}$ into $\{\hg\mkl\}_{m\in\A_k, l\in[L]}$ accordingly.

    \item Stage 4: Each UE feeds the quantized path gains back to every associated AP. For example, UE $k$ feeds $(\hg\mkI, \ldots, \hg\mkL)$ back to AP $m$ for every $m\in\A_k$.

    \item Stage 5: Each AP constructs the quantized channels and serves its own UE group with ZF beamforming.
\end{itemize}

\subsection{Solution of Problem \PAP{3}}
In the problem \PAP{3}, the task entails determining both the UE-AP association and the allocation of feedback bits. However, directly solving \PAP{3} proves challenging due to its inclusion of UE selection, rendering it a mixed-integer problem recognized as NP-hard. Consequently, each AP tackles \PAP{3} through a two-step approach. Initially, each AP identifies the UE-AP associations assuming an equal allocation of feedback resources (e.g., $B_m$) among associated UEs, with the feedback bits allocated per UE being evenly distributed across all communication paths, i.e.,
\begin{align}
  \PAP{3a}~~~
  \underset{\{\U_m \subset\Uo_m\}_{m\in[M]}}
  {\mathrm{maximize}}
    ~~&\Rsum'
       ~\Big\vert_{b\mkl=\frac{B_m}{\vert\U_m\vert L},~ \forall m,k,l}
    \NNL
    \mathrm{subject~to}~~~~~
    &\eqref{cconstraint0}.\NN
\end{align}
Then, in the second step, each AP finds the feedback bit allocation/splitting with the UE-AP association obtained from the problem \PAP{3a} as follows:
\begin{align}
  \PAP{3b}~~~
  \underset{\{b\mkl\}_{m\in[M], k\in[K], l\in[L]}}
  {\mathrm{maximize}}
    ~~&\Rsum'
    \NNL
    \mathrm{subject~to}~~~~~~~~
    &\eqref{cconstraint1}, \eqref{cconstraint2}, \eqref{cconstraint3}.\NN
\end{align}

\subsubsection{Solution of Problem \PAP{3a}}
The problem \PAP{3a} is also a mixed-integer problem, making it inherently difficult to solve. Consequently, we opt for a suboptimal approach, wherein each AP independently finds each of $\{\U_1, \ldots, \U_M\}$ from the initial UE-AP association, by relaxing (or ignoring) other UE groups' cardinality constraints. For example, to determine $\U_1$, each AP finds the best $\U_1\subset\Uo_1$ such that $\vert\U_1\vert = \min(\Uo_1, N_s)$ by simply setting $\U_m=\Uo_m$ for all $m\in\{2, \ldots, M\}$. In this way, the AP finds $\U_m$ for all $m\in[M]$ as follows.
\begin{itemize}
\item If $\vert\Uo_m\vert \le N_s$, set $\U_m = \Uo_m$.

\item If $\vert\Uo_m\vert > N_s$, choose the best $\U_m\subset \Uo_m$ such that $\vert\U_m\vert=N_s$, which maximizes $\Rsum'\big\vert_{ b\mkl=\frac{B_m}{\vert\U_m\vert L},~ \forall m,k,l}$, fixing other UE groups as the initial UE groups, i.e., $\U_j=\Uo_j$ for all $j\in[M]\setminus\{m\}$.
\end{itemize}

With a slight abuse of notation, we denote by $\Rsum^\circ$  the achievable sum rate with the initial UE-AP association and the equal feedback bit allocation/splitting oblivious to each AP's stream number constraint (i.e., each AP's UE group cardinality constraint) as follows.
\begin{align}
  \Rsum^\circ
  \triangleq \Rsum'(\Uo_1, \ldots, \Uo_M)\big\vert_{b\mkl=\frac{B_m}{\vert\Uo_m\vert L},~ \forall m,k,l}.
\end{align}
The procedure to solve the problem \PAP{3a} can be summarized as in Algorithm 1.

\begin{algorithm}[t]
\caption{Proposed User-Centric UE-AP Association}\label{algorithm1}
\begin{algorithmic}[1]
\State \textbf{Input:} The initial UE-AP association $(\Uo_1, \ldots, \Uo_M)$
\State \textbf{Output:} The revised UE-AP association $(\U_1, \ldots, \U_M)$
\ForAll{$m\in[M]$}
    \If{$\vert\Uo_m\vert \le N_s$}
        \State $\U_m=\Uo_m$.
    \Else {~~(when $|\Uo_m|>N_s$)}
        \State Randomly choose $\U_m\subset\Uo_m$ such that $\vert \U_m\vert =N_s$.
        \ForAll{$\U_m'\subset\Uo_m$ such that $\vert \U_m'\vert =N_s$}
            \If{$\Rsum^\circ \big\vert_{\Uo_m=\U'_m}
                \ge \Rsum^\circ \big\vert_{\Uo_m=\U_m}$}
                \State $\U_m = \U'_m$.
            \EndIf
        \EndFor
    \EndIf
\EndFor
\end{algorithmic}
\end{algorithm}

\subsubsection{Solution of Problem \PAP{3b}\label{PAP3b_solution}}
To solve the problem \PAP{3b}, we initially determine the equivalents of the objective function as follows:
\begin{align}
&\underset{\{b\mkl\}_{\forall m,k,l}}{\mathrm{maximize}}
   ~~\Rsum'~~
\stackrel{(a)}{\longleftrightarrow}~~
  \underset{\{b\mkl\}_{\forall m,k,l}}{\mathrm{minimize}}
   ~~\tsuml_{k\in[K]}\I_k^{\SGI\downarrow}~\NNL
&\stackrel{(b)}{\longleftrightarrow}
\underset{\{b\mkl\}_{\forall m,k,l}}{\mathrm{minimize}}~~
  \tsuml_{k\in[K]}
  \tsuml_{\substack{j\in[K]    \\   j\ne k}}
  \tsuml_{\substack{m\in\A_j   \\   m\in\A_k}}
  \tsuml_{l\in[L]}
  \pmk\beta_{m,k}\kappa\mkl 2^{-b\mkl}\NNL
&\stackrel{(c)}{\longleftrightarrow}
\underset{\{b\mkl\}_{\forall m,k,l}}{\mathrm{minimize}}
  \tsuml_{m\in[M]}
  \tsuml_{k\in\U_m}
  \tsuml_{l\in[L]}
  \Pmk\beta_{m,k}\kappa\mkl 2^{-b\mkl},\NN
\end{align}
where $\Pmk\triangleq (\vert\U_m\vert -1)\pmk$\KJcancel{ is introduced for notational simplicity}. In the expression above, the equivalence $(a)$ is from \eqref{eqn:Rsum'} with \eqref{eqn:D_k}--\eqref{eqn:I_k_IGI} (\KJcancel{refer to}\magenta{see} Observation 1),\KJcancel{ and} \magenta{the equivalence} $(b)$ is from \eqref{eqn:I_k_SGI}\KJcancel{. Also}, \magenta{and} the equivalence $(c)$ is obtained after a simple manipulation by representing the AP clusters with the UE groups. Thus, we obtain the equivalent problem of \PAP{3b} as follows:
\begin{align}
\PAP{3b$'$}&~~
\underset{\{b\mkl\}_{\forall m,k,l}}{\mathrm{minimize}}~
  \tsuml_{m\in[M]}
  \tsuml_{k\in\U_m}
  \tsuml_{l\in[L]}
  \Pmk\beta_{m,k}\kappa\mkl 2^{-b\mkl} \NNL
\mathrm{s.t.}\quad&
    \tsuml_{k\in \U_m} \tsuml_{l\in[L]}b\mkl=B_m,
        \quad
        \forall m\in[M] \label{ccconstraint1}\\
    & ~b\mkl\ge 0,
        \quad
        \forall m\in[M]~~\forall k\in\U_m~~\forall l\in[L]\\
    & ~b\mkl=0,
        \quad
        \forall m\in[M]~~\forall k\not\in\U_m~~\forall l\in[L].
\end{align}

Since the objective function of the problem \PAP{3b$'$} is a convex function of $b\mkl$ for any $(m, k, l)$ as well as the constraint functions, the problem \PAP{3b$'$} becomes a convex problem. Thus, to solve the problem \PAP{3b$'$}, we construct a Lagrangian $\mathcal{L}$ as follows:
\begin{align}
\mathcal{L}
  &\teq
  \tsuml_{m\in[M]}
  \tsuml_{k\in\U_m}
  \tsuml_{l\in[L]}
  \Pmk\beta_{m,k}\kappa\mkl 2^{-b\mkl}  \NNL
  &\qquad - \tsuml_{m\in[M]}
  \tsuml_{k\in\U_m}
  \tsuml_{l\in[L]}
  \mu\mkl b\mkl
  \NNL
  &\qquad
  +\tsuml_{m\in[M]}\lambda_m
  \Big(\tsuml_{k\in\U_m}\tsuml_{l\in[L]} b\mkl-B_m\Big),
\end{align}
where $\{\mu\mkl\}_{m\in[M], k\in\U_m, l\in[L]}$ and $\{\lambda_m\}_{m\in[M]}$ are the Lagrangian multipliers. Then, from KKT conditions, which are the optimality conditions in our case, we have $\mu\mkl b\mkl=0$ for any $(m, k, l)$. Also, the optimal solution denoted by $\{b\mkls\}$ is obtained at $\partial \mathcal{L}(b\mkls) / \partial b\mkl=0$ whenever $b\mkls>0$, or $b\mkls=0$ otherwise. To make $\partial \mathcal{L}/\partial b\mkl=0$, it should be satisfied that $b\mkl = \log_2(\ln2 \cdot \Pmk \beta_{m,k} \kappa\mkl / \lambda_m )$. Then, owing to the KKT conditions, we obtain the optimal feedback bit allocation/splitting as follows:
\begin{align}
b\mkls=
\begin{cases}
    \big[\log_2(\beta_{m,k}\kappa\mkl) - \lambda'_m\big]^+,
        &\textrm{~if~}~k\in\U_m\\
    0,  &\textrm{~otherwise,}
\end{cases}
\label{feedback_bit_solution}
\end{align}
where $\lambda'_m \triangleq \log_2(\lambda_m/(\ln2\cdot \Pmk))$ is a constant that should be numerically found to satisfy the feedback link sharing constraint \eqref{ccconstraint1}, i.e., $\sum_{k\in\U_m}\sum_{l\in[L]}b\mkls = B_m$.

\begin{remark}
The feedback bit allocation/splitting given in \eqref{feedback_bit_solution} is only based on the long-term varying elements as we mentioned in Remark 1. From \eqref{feedback_bit_solution}, we can observe that AP $m$ allocates its feedback link budget to a total of $\vert\U_m\vert\cdot L$ dominant paths of its associated UEs. In this case, the allocated feedback bits increases as the large-scale fading coefficient (e.g., $\beta\mk$) increases and the dominance factor (e.g., $\kappa\mkl$) increases.
\end{remark}

\subsection{Solution of Problem \PUE{2}\label{PUE2_solution}}
\Blue
Note that solving $\PAP{3}$ allows the AP to obtain both the feedback bit allocation to each associated UE and the feedback bit splitting for each path of the corresponding UE. Without solving $\PUE{2}$ at UE $k$, AP $m$ needs to inform UE $k$ of $L$ scalar values $(b\mkI, \dots, b\mkL)$ for feedback bit allocation. In contrast, in our scheme, AP $m$ provides only a single scalar value $b\mk$, and UE $k$ determines the feedback splitting $(b\mkI, \dots, b\mkL)$ by solving $\PUE{2}$. In this way, we can reduce the downlink control channel overhead.\KJcancel{ When the UE solves $\PUE{2}$ based on $b\mk$, it solves the same convex problem as the AP solves $\PAP{3}$, so the results are consistent with those of $\PAP{3}$.}

\Black
The problem \PUE{2} can be solved similarly to the problem \PAP{3b$'$}, so we briefly explain its solution. We can find the equivalent objective function of \PUE{2} as follows:
\begin{align}
&\underset{\{b\mkl\}_{m\in\A_k, l\in[L]}}
  {\mathrm{maximize}} ~~\R_k' \NNL
&\stackrel{(a)}{\longleftrightarrow}
  \underset{\{b\mkl\}_{m\in\A_k, l\in[L]}}
  {\mathrm{minimize}}~~
    \tsuml\limits_{\substack{j\in[K]    \\   j\ne k}}
    \tsuml\limits_{\substack{m\in\A_j   \\   m\in\A_k}}
    \tsuml\limits_{l\in[L]}
        \pmk\beta_{m,k}
        \kappa\mkl 2^{-b\mkl} \NNL
&\stackrel{(b)}{\longleftrightarrow}
  \underset{\{b\mkl\}_{m\in\A_k, l\in[L]}}
  {\mathrm{minimize}}~~
    \tsuml\limits_{\substack{m\in\A_k}}
    \tsuml\limits_{l\in[L]}
        \Pmk\beta_{m,k}
        \kappa\mkl 2^{-b\mkl},\NN
\end{align}
where the equivalence $(a)$ is from \eqref{eqn:Rk'}\KJcancel{\eqref{eqn:Rk'_full}}, while $(b)$ is obtained
from the identity for any function $f(m,k)$ given by
\begin{align}
 \tsuml\limits_{\substack{j\in[K]    \\   j\ne k}}
 \tsuml\limits_{\substack{m\in\A_j   \\   m\in\A_k}}
 f(m,k)
 =
 \tsuml\limits_{\substack{m\in\A_k}} (\vert\U_m\vert-1)f(m,k).
\end{align}

Consequently, the problem \PUE{2} is equivalent to the following problem:
\begin{align}
\PUE{2$'$}~
  &\underset{\{b\mkl\}_{m\in\A_k, l\in[L]}}
  {\mathrm{minimize}}~~
\tsuml\limits_{\substack{m\in\A_k}}
    \tsuml\limits_{l\in[L]}
        \Pmk\beta_{m,k}
        \kappa\mkl 2^{-b\mkl}
 \NNL
    &\mathrm{s.t.}~~
    \textstyle\sum_{l\in[L]} b\mkl =b_{m,k}, \qquad \forall m\in\A_k\label{cconstraint_k1}\\
    &\qquad b\mkl\ge 0,  \qquad\forall m\in\A_k ~~\forall l\in[L].
\end{align}
As the problem \PUE{2$'$} is the convex problem, we obtain the optimal solution as follows:
\begin{align}
b\mkls=
\begin{cases}
    \big[\log_2(\kappa\mkl) - \lambda''\mk\big]^+,
        &\textrm{~if~}~m\in\A_k\\
    0,  &\textrm{~otherwise,}
\end{cases}
\label{feedback_bit_splitting_solution}
\end{align}
where $\lambda''\mk$ is a constant that should be numerically found to satisfy the feedback bit splitting constraint \eqref{cconstraint_k1}, i.e., $\sum_{l\in[L]}b\mkls = b\mk$. Thus, once the feedback bits $b\mk$ are allocated from AP $m$, UE $k$ obtains $\{b\mkl\}_{l=1}^L$ according to \eqref{feedback_bit_splitting_solution}. In \eqref{feedback_bit_splitting_solution}, we can observe that the optimal feedback bit splitting is dependent only on its own dominance factors, as we stated in Remark 2, and each UE allocates more feedback bits to the dominant path with a larger path gain.

\begin{remark}
The solution \eqref{feedback_bit_splitting_solution} becomes equivalent to the solution \eqref{feedback_bit_solution} by introducing a new constant
$
  \lambda\mk'''
  \triangleq \lambda\mk''+\log_2(\beta\mk)
$
into \eqref{feedback_bit_splitting_solution}, replacing $\lambda\mk''$. In this case, we have $\lambda\mk'''=\lambda_m'$ for all $k\in\U_m$, where $\lambda_m'$ is a constant used in \eqref{feedback_bit_solution}, which validates the composite problem \PAP{3}.
\end{remark}

\Blue\KJcancel{
\subsection{Each UE's Computational Burden}
In this subsection, we analyze the computational burden on each UE. At each UE, the computational burden depends on the number of associated APs (e.g., $\vert\A_k\vert$ for UE $k$) and the number of paths (i.e., $L$). As mentioned in Section \ref{procedure}, UE $k$ receives a single scalar value from each associated AP (e.g., $b\mk$ for $m\in\A_k$), resulting in a total of $\vert\A_k\vert$ scalar values. Based on the $b\mk$ for all $m\in\A_k$, the computational burden of each UE can be divided into two main factors: the complexity of solving the problem $\PUE{2$'$}$ and the complexity of quantizing the path gains.

\emph{1) Complexity Analysis of Problem \PUE{2$'$}: }
Firstly, UE $k$ finds the solution given in \eqref{feedback_bit_splitting_solution} $\vert\A_k\vert$ times to determine $b\mkl$ for all $m\in\A_k$ and $l\in[L]$. The problem $\PUE{2$'$}$ is convex in $b\mkl$, and the solution \eqref{feedback_bit_splitting_solution} provides a closed-form expression for the optimal $b\mkl$. This can be efficiently computed using basic arithmetic and logarithmic operations, allowing for an iterative solution with a complexity of $\mathcal{O}(\vert\A_k\vert\cdot L)$.

\emph{2) Complexity Analysis of Rate-Distortion Quantizer: }
Secondly, UE $k$ quantizes each path gain (e.g., $g\mkl$) according to the feedback bits $b\mkl$. In Section \ref{identification}, each path gain is quantized independently for its real and imaginary parts, resulting in $2L$ scalar quantizations per associated AP, achieved through a total of $2L$ product codebooks.

Rate-distortion quantization involves minimizing the distortion for a given feedback bit budget, often implemented using techniques like the Lloyd-Max algorithm. If Lloyd-Max quantization is used for scalar quantization of each real and imaginary part of $g\mkl$, the complexity of each quantization is approximately $\mathcal{O}(n\log{n})$, where $n$ represents the number of iterations needed for convergence in the Lloyd-Max algorithm. Thus, the complexity of the rate-distortion quantizer for all real and imaginary parts across all associated APs and paths becomes $\mathcal{O}(\vert\A_k\vert\cdot L\cdot n\log{n})$.

Finally, combining both factors, the overall computational burden on each UE, incorporating both $\PUE{2$'$}$ and rate-distortion quantizer, is given by $\mathcal{O}(\vert\A_k\vert\cdot L\cdot n\log{n})$. Although UEs need to perform these quantization computations for the determined feedback bits at intervals consistent with the short-term varying elements, this level of complexity remains manageable due to the scalar nature of quantization applied independently to each real and imaginary component, allowing relatively fast convergence within a small number $n$ of iterations. Additionally, modern UE hardware is typically equipped to support such problem-solving and quantization processes within these short intervals, benefitting from optimizations in processing power and energy efficiency.
}

\Black
\section{Numerical Results\label{numericalResults}}
To evaluate our proposed scheme, we mainly consider a \blue{CF-mMIMO}\KJcancel{network} scenario, where forty APs equipped with eight antennas each and twenty single-antenna UEs, i.e., $(M, N, K)=(40, 8, 20)$, are uniformly distributed in a square area of $1\times 1$ km$^2$ with a wrap-around setup \cite{Ozdogan2019}. \KJcancel{An example of our \blue{CF-mMIMO}\KJcancel{network} scenario is illustrated in Fig. \ref{fig2}. }We assume that each AP utilizes at most four streams (i.e., $N_s=4$), while the simulation result for varying $N_s$ will also be provided. We also assume that each AP has a ten-bit uplink feedback budget, i.e., $B_1 = \dots = B_M = 10$. Performance evaluation configurations can be adjusted as necessary and will be specified as required.

We employ the Saleh-Valenzuela channel model with a center frequency of $3.5$ GHz and a bandwidth of $100$ MHz, which is widely used to model the downlink of sub-6 GHz bands for 5G NR systems \cite{Tech2024}. The large-scale fading coefficients (e.g., $\beta\mk$) are calculated based on the COST 231 Walfisch-Ikegami model \cite{Tech2022} with the AP height of $12.5$ m and the UE height of $1.5$ m. Thus, the large-scale fading coefficients are given, in decibels, by
\begin{align}\label{pathLoss}
    \beta_{m,k}=-30.18-26\log_{10}(d_{m,k}),
\end{align}
where $d_{m,k}$ is the distance between AP $m$ and UE $k$, whose unit is meter.

We assume that each channel comprises four dominant paths (i.e., $L=4$). To model $L$ dominance factors at each channel, we independently generate $(L-1)$ points in the interval $[0, 1]$ from a uniform distribution and obtain $L$ subintervals. The sequence of the first to the $L$th dominance factors is then obtained from the sorted lengths of the $L$ subintervals, arranged in descending order. This ensures that their sum equals one, with the $L$th channel component having the smallest dominance factor. Meanwhile, the angular spread for AoDs is set to $20^\circ$. \blue{Details of our CF-mMIMO scenario are summarized in Table \ref{table:parameters}, along with a comparison to other environments used in previous works \cite{Ngo2017, Buzzi2020, Abdallah2020, Ozdogan2019}.}

\begin{table}\Blue
\centering
\caption{The Comparison of Various CF-mMIMO Scenarios}
\label{table:parameters}
\begin{tabular}{c|ccccccc}
\hline\hline
    \multirow{2}{*}{CF-mMIMO scenarios}
    & \multicolumn{7}{c}{Parameters}\\
    \cline{2-8}
      & $M$ & $N$ & $MN$ & $K$ & $N_s$ & $L$ & $B$ \\
\hline\hline
 Our scenario &
    40 & 8 & 320 & 20 & 4 & 4 & 10\\
\hline
 Scenarios in \cite{Ngo2017}, \cite{Ozdogan2019} &
     100 & 1 & 100 & 40 & - & - & -\\
\hline
 Scenario in \cite{Buzzi2020} &
     80 & 6 & 480 & 15 & - & - & -\\
\hline
 Scenario in \cite{Abdallah2020} &
     10 & 32 & 320 & 20 & - & 4 & -\\
\hline\hline
\end{tabular}
\vspace{-0.4cm}
\end{table}

\KJcancel{\begin{figure}[!t]
\centering
\includegraphics[width=0.8\columnwidth]{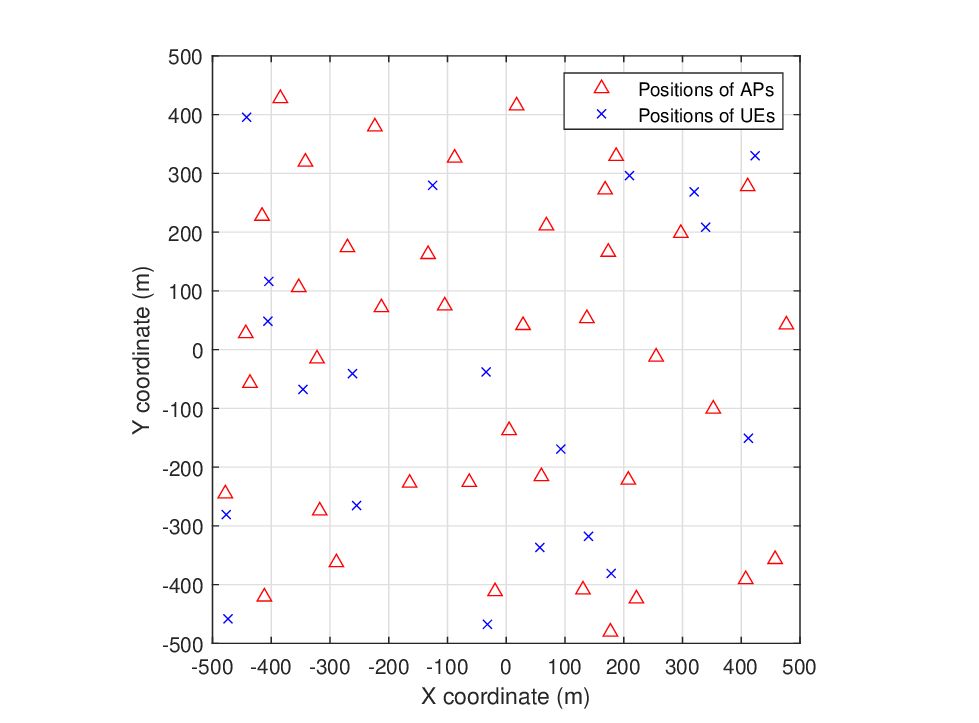}
\caption{An illustration of our \blue{CF-mMIMO} \KJcancel{network }scenario\blue{, where}
40 APs and 20 UEs are uniformly distributed in a $1\times1$ km$^2$ area.}
\label{fig2}
\vspace{-0.4cm}
\end{figure}}

For comparison, we consider three user-centric association methods and three feedback bit allocation methods separately, resulting in a total of nine association and feedback bit allocation schemes. The UE-AP association methods start with a simple initial UE-AP association, where each UE selects the nearest $\lceil M/2\rceil$ APs, which have the largest large-scale fading coefficients, i.e., for every $k\in[K]$, the initial AP cluster $\Ao_k$ satisfies that $\vert\Ao_k\vert=\lceil M/2\rceil$ and
\begin{align}
  \min_{m\in\Ao_k} \beta\mk \ge \max_{m\in[M]\setminus\Ao_k} \beta\mk.
\end{align}
Then, UE groups are determined from the initial association in one of three methods as follows:
\begin{itemize}
\item Proposed association: UE groups are determined from the initial UE groups according to Algorithm 1.

\item Beta association: Each UE group is formed by choosing $N_s$ UEs with the largest large-scale fading coefficients.

\item Random association: Each UE group is determined by randomly choosing $N_s$ UEs in each initial UE group.
\end{itemize}
Also, we consider three feedback bit allocation methods as follows:
\begin{itemize}
\item Perfect CSI: Each AP perfectly knows the CSI, which corresponds to infinity feedback bit allocation.

\item Proposed feedback bit allocation: The feedback bits are allocated to the associated UEs according to \eqref{feedback_bit_solution}.

\item Equal feedback bit allocation: The feedback bits are equally allocated to the associated UEs for each AP.
\end{itemize}

\begin{figure}[t!]
   \centering
   \subfigure[]
   {\includegraphics[width=0.8\columnwidth]{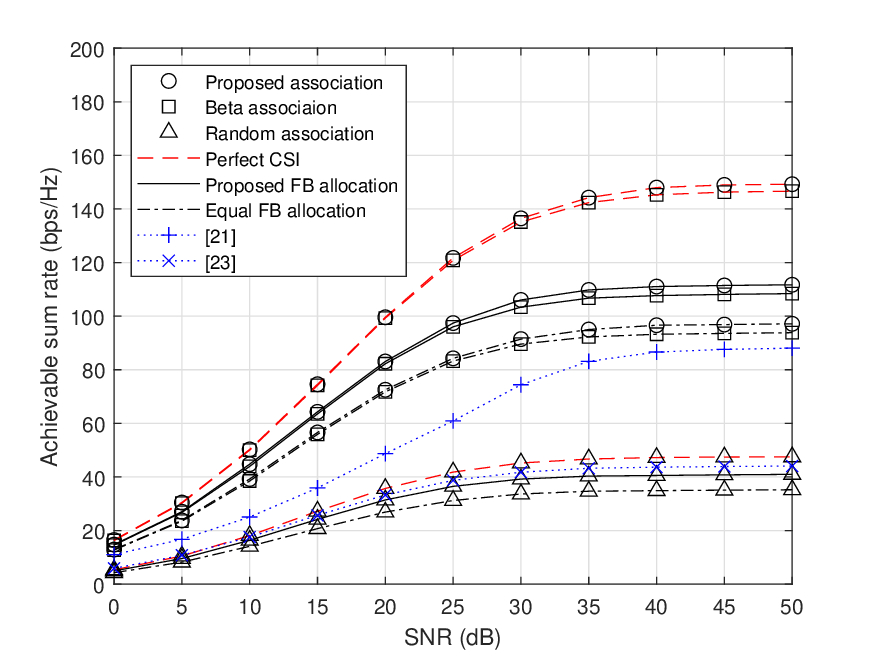}}
   \subfigure[]
   {\includegraphics[width=0.8\columnwidth]{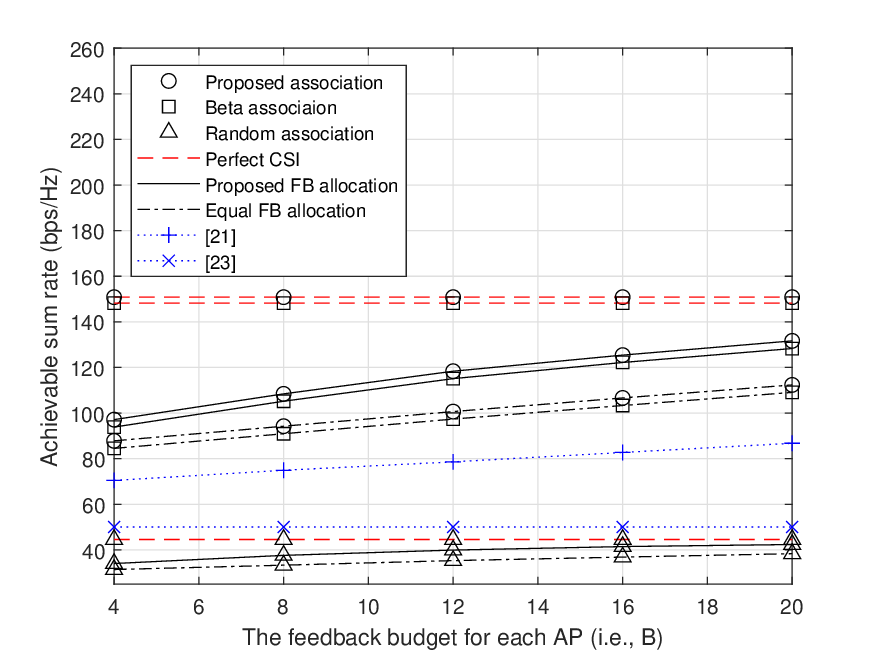}}
   \vspace{-0.25cm}
   \caption{Achievable sum rates of various schemes \blue{in our CF-mMIMO scenario} with respect to (a) the transmit SNR and (b) each AP's feedback budget.}
   \label{fig4}
   \vspace{-0.4cm}
\end{figure}

In addition to the aforementioned schemes, we consider \blue{two} more reference schemes as follows.
\begin{itemize}
\item The reference scheme \cite{Kim2020}: The CPU selects for each UE a subset of total $ML$ dominant paths from all APs that maximizes the signal-to-leakage-and-noise ratio (SLNR). Each UE then feeds back a quantized version of a vector comprising the selected path gains. Based on the feedback information, each AP serves UEs with the precoding strategy that maximizes the SLNR. Although this scheme is not directly applicable to our system model with feedback link sharing, we assume that each UE is assigned a feedback size of $B$ bits, the same as the uplink feedback budget. Also, we assume that eight dominant paths are selected for each UE from the total paths.

\item The reference scheme \cite{Abdallah2020}: Each AP utilizes the angle-based ZF beamforming vector, obtained solely from long-term varying elements. Thus, each UE does not need to provide any feedback. Our proposed association method is adopted with the simple initial UE-AP association.

\end{itemize}

\KJcancel{
\begin{figure}[!t]\Red
    \centering
    \includegraphics[width=0.7\columnwidth]{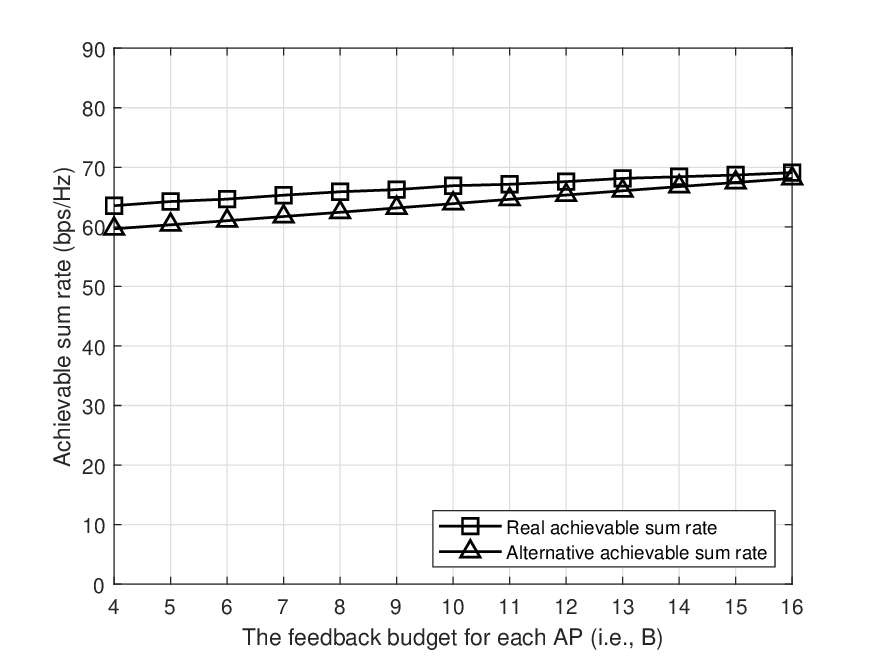}
    \caption{The alternative sum rate $\Rsum'$ and the actual sum rate $\Rsum$ with respect to each AP's feedback budget. \blue{\KJcancelc{Our alternative sum rate $\Rsum'$ effectively reflects the actual sum rate $\Rsum$.}}}
    \label{fig3}
    \vspace{-0.4cm}
\end{figure}}

\KJcancel{
Since our proposed solution relies on the \blue{\KJcancelc{alternative}} sum rate $\Rsum'$ (i.e., \eqref{eqn:Rsum'}), we first compare in Fig. \ref{fig3} the \blue{\KJcancelc{alternative}} sum rate $\Rsum'$ with the average sum rate $\E[\Rsum]$ in \blue{\KJcancelc{our CF-mMIMO}} network scenario, while varying each AP's feedback budget. This comparison serves to gauge the accuracy of our approximation \blue{\KJcancelc{and the lower bound}}. Specifically, we set  each AP's transmit SNR to 20 dB, and employ the proposed association with an equal allocation of feedback. Fig. \ref{fig3} illustrates that the \blue{\KJcancelc{alternative}} sum rate $\Rsum'$ effectively mirrors the actual sum rate $\Rsum$.}


\begin{figure*}[t!]
    \centering
    \subfigure[]
    {\includegraphics[width=.32\textwidth]{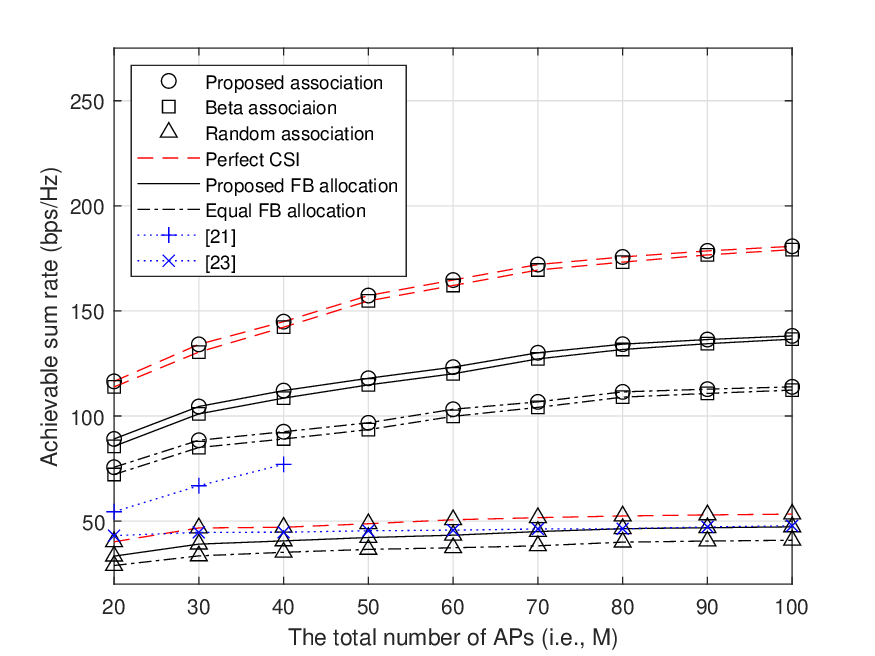}}
    \subfigure[]
    {\includegraphics[width=.32\textwidth]{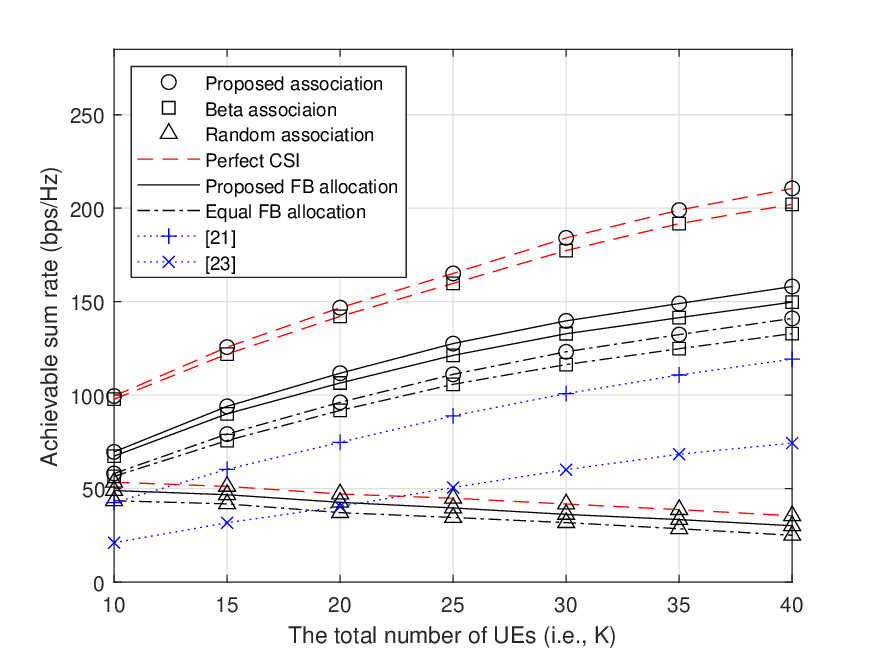}}
    \subfigure[]
    {\includegraphics[width=.32\textwidth]{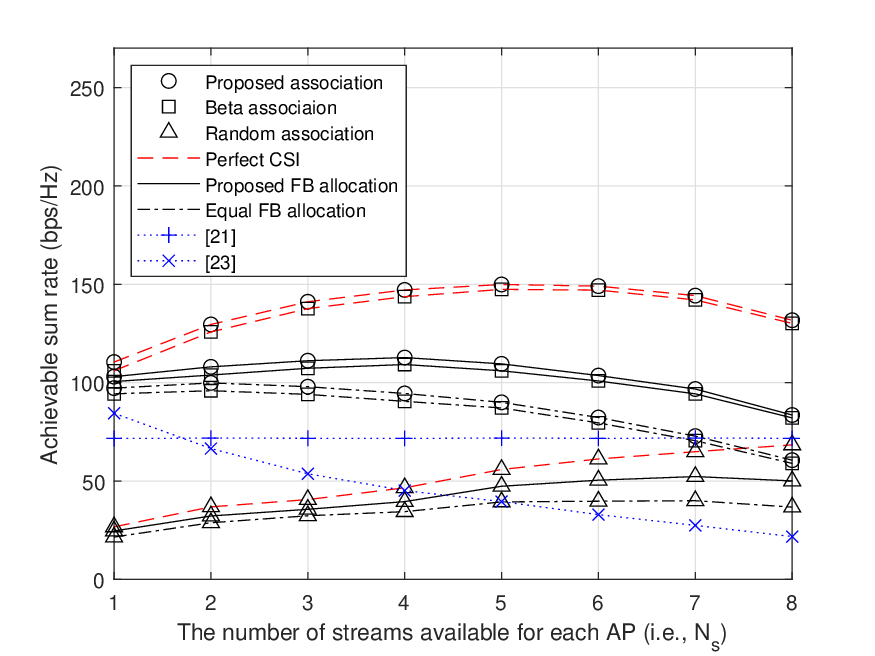}}
    \vspace{-0.25cm}
    \caption{Achievable sum rates of various schemes \blue{in our CF-mMIMO scenario} with respect to (a) the total number of APs, (b) the total number of UEs, (c) the maximum number of streams available at each AP. \blue{\KJcancel{Our proposed association and feedback bit allocation schemes enhance the performance of conventional schemes.}}}
    \label{fig5}
    \vspace{-0.4cm}
\end{figure*}

\blue{In Fig. \ref{fig4} and Fig. \ref{fig5}, we evaluate the achievable sum rate of our scheme in our CF-mMIMO scenario by varying various factors: Fig. \ref{fig4}(a) the transmit SNR, Fig. \ref{fig4}(b) the feedback budget at each AP, Fig. \ref{fig5}(a) the number of APs, Fig. \ref{fig5}(b) the number of UEs, and Fig. \ref{fig5}(c) the number of streams available at each AP.}
In Fig. \ref{fig4}(a), we show the achievable sum rates of various schemes with respect to the transmit SNR. We can observe in Fig. \ref{fig4}(a) that the sum rate with each scheme increases as the transmit SNR increases, but is saturated in the high SNR region. This is because the IGI remains in each UE's received signal even in the perfect CSI case, while not only the IGI but also the SGI due to quantization errors persist with the received signal in the limited feedback environment.
\blue{Additionally, as observed in Fig. \ref{fig4}(a), our proposed association method outperforms the beta association method in limited feedback environments because it is based on a unified measure that accounts for both the long-term varying elements and the performance degradation caused by imperfect CSI. In addition, while the SGI term vanishes under perfect CSI conditions, the effects of IGI are still reflected, demonstrating that our proposed method surpasses the performance of the beta association method.}
Furthermore, our proposed feedback bit allocation method enhances the performance of the equal feedback bit allocation method for each association method. Moreover, the combination of our proposed association and feedback bit allocation surpasses the performance of the reference schemes \cite{Kim2020} and \cite{Abdallah2020}. These observations underscore the effectiveness of our proposed user-centric association and feedback bit allocation in improving the performance of conventional schemes.

In Fig. \ref{fig4}(b), we compare the achievable sum rates of various schemes with respect to the uplink feedback budget at each AP, i.e., $B$, when each AP's transmit SNR is 50 dB.\footnote{The 50 dB SNR may not be considered large when taking into account large-scale fading. Note that according to \eqref{pathLoss}, the average receive SNR at a distance of 10 m is approximately $-6$ dB.} As shown in Fig. \ref{fig4}(b), the sum rate for each scheme increases as the uplink feedback budget increases, allowing each AP to utilize more precise ZF beamforming vectors. Notably, our proposed association method consistently outperforms the other two association methods, regardless of the uplink feedback budget. Additionally, our devised feedback bit allocation method enhances the performance compared to the equal feedback bit allocation method, across varying uplink feedback budgets and irrespective of the specific association method used.

In Fig. \ref{fig5}(a), we compare the achievable sum rates of various schemes, varying the total number of APs (i.e., $M$) while maintaining each AP's transmit SNR at 50 dB.\footnote{Note that the reference scheme \cite{Kim2020} is evaluated with a maximum of 40 APs due to limited computing power; it has computational complexity proportional to $ML$, leading to prohibitively high computational complexity as the number of APs grows.} As illustrated in Fig. \ref{fig5}(a), we note a positive correlation between the achievable sum rate and the number of APs in the network. This relationship stems from the enhanced AP selection diversity gain afforded by a greater density of APs within the same coverage area, benefiting each UE.


In Fig. \ref{fig5}(b), we compare the achievable sum rates of various schemes, adjusting the total number of UEs (i.e., $K$) with each AP's transmit SNR at 50 dB. Fig. \ref{fig5}(b) reveals that both our proposed association method and the beta association methods exhibit increasing sum rates as the number of UEs rises, regardless of the feedback allocation method employed. Conversely, the sum rate for the random association decreases with the growth in total UEs. This decline can be attributed to the random selection process, which becomes more likely to pick UEs experiencing severe path losses or shadowing effects as the pool of UEs expands. \blue{In addition}, we observe consistent trends wherein our proposed association method and feedback bit allocation scheme consistently outperform their counterparts. This observation underscores the efficacy of our proposed scheme in enhancing the performance of conventional approaches.

In Fig. \ref{fig5}(c), we compare the achievable sum rates of various schemes, while varying the number of streams available at each AP (i.e., $N_s$), and maintaining a transmit SNR of 50 dB. This figure reveals that the sum rates reach their peak at specific stream counts for both the proposed association and beta association methods. For instance, optimal performance in our proposed association is observed at $N_s=5$ with the perfect CSI and \blue{$N_s=4$} with the feedback bit allocation methods. Notably, when $N_s$ equals to the total number of streams ($N_s=N =8$), the achievable sum rates rather decreases. Increasing $N_s$ allows each AP to accommodate more UEs through multiuser beamforming, potentially enhancing multiplexing gain. However, this expansion comes with a trade-off: as more UEs share the uplink feedback budget, the AP's CSI knowledge becomes less precise, leading to performance degradation due to imprecise beamforming vectors. These findings underscore the importance of carefully selecting the maximum stream count, taking into account various influencing factors.

\KJcancel{
\begin{figure}[!t]\Red
    \centering
    \includegraphics[width=0.7\columnwidth]{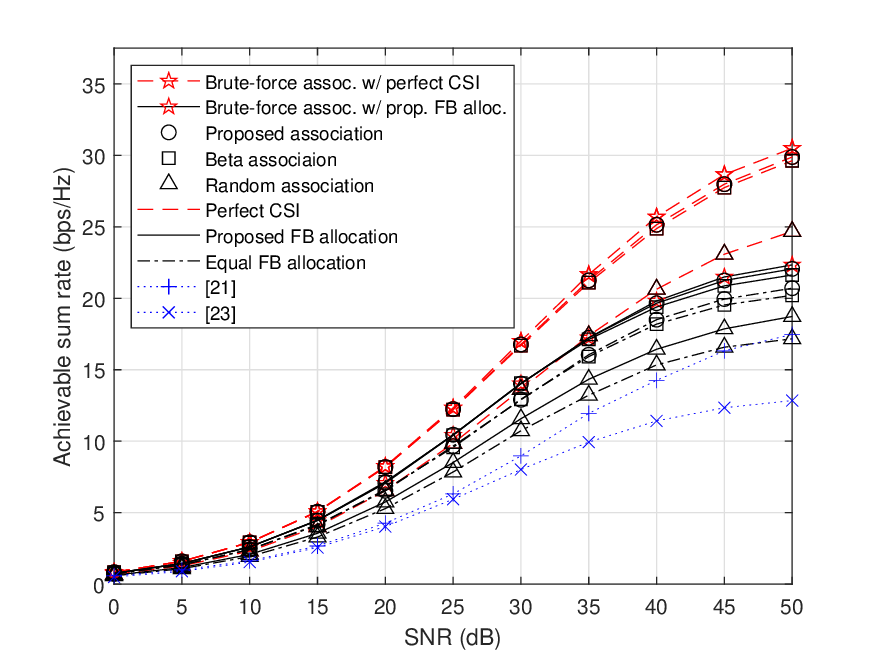}
    \caption{\blue{\KJcancelc{Achievable sum rates of various schemes with respect to the transmit SNR for the network parameters $(M, N, K) = (8, 4, 4)$. Our proposed association and feedback bit allocation schemes enhance the performance of conventional schemes.}}}
    \label{fig6}
    \vspace{-0.4cm}
\end{figure}}

\blue{\KJcancel{To further evaluate our proposed association and feedback bit allocation schemes, we also consider the brute-force type reference schemes for comparison.}}
\blue{\KJcancel{For the performance evaluation, we consider a smaller network size with parameters of $(M, N, K) = (8, 4, 4)$, as the brute-force type reference schemes are not feasible under our CF-mMIMO scenario summarized in Table \ref{table:parameters}. The brute-force type reference schemes are as follows.}}
\KJcancel{
\begin{itemize}
\item Brute-force association with perfect CSI: The optimal solution of the problem \PAP{3} when each AP has perfect CSI. In this scheme, each AP finds the optimal association using brute-force search, with all feedback sizes set to infinity, i.e., $b\mkl=\infty$ for all $m, k, l$.

\item Brute-force association with proposed feedback bit allocation: Each AP finds the optimal association using brute-force search, adopting our proposed feedback bit allocation, i.e., \eqref{feedback_bit_solution}.
\end{itemize}}
\blue{\KJcancel{Fig. \ref{fig6} illustrates the achievable sum rates of various schemes with respect to the transmit SNR under the network parameter setting of $(M, N, K) = (8, 4, 4)$. As depicted, while the brute-force type scheme demonstrates slightly higher performance in both perfect CSI and limited feedback environments, our proposed association and feedback bit allocation schemes exhibit competitive performance compared to the brute-force type schemes with significantly lower computing power.}} \KJcancel{Notably, similar trends to those observed in Fig. \ref{fig4}(a) are apparent.}

\section{Conclusion\label{conclusion}}\Blue
In this paper, we introduced a novel framework to optimize user-centric UE-AP association and feedback bit allocation for FDD CF-mMIMO systems operating under limited feedback environments. Using the Saleh-Valenzuela channel model with varying average path power, we identified critical feedback information and developed a tailored codebook structure, integrating these into a unified optimization framework with an alternative sum rate approximation to enable tractable solutions. Our proposed protocol effectively refines associations and allocates feedback bits across UEs and their multiple paths based on long-term channel characteristics. Simulation results demonstrated the superior performance of our scheme compared to existing high-complexity methods \magenta{in our CF-mMIMO scenario, which}\KJcancel{particularly in large-size network scenarios that} closely reflect\magenta{s} practical CF-mMIMO deployments. \KJcancel{These findings emphasize the importance of optimizing UE-AP association and feedback bit allocation to maximize system performance and scalability. }This work provides a solid foundation for future research to explore adaptive schemes that address dynamic channel variations and feedback budget constraints, bridging the gap between theoretical advancements and real-world applications in next-generation wireless systems.

\Black
\appendices
\section{Proof of Lemma 3}\Blue
To prove Lemma 3, we use the fact that the variance of the sum of independent random variables equals the sum of their variances. This can be summarized in the following remark.
\begin{remark}
Let $X_1, \ldots, X_n$ be mutually independent complex random variables with
$\E[X_1] = \ldots = \E[X_n] = 0$. \KJcancel{Then, w}We have
\begin{align}
  \E [ \vert X_1 + \ldots + X_n \vert^2]
  =
  \E [\vert X_1 \vert^2] + \ldots + \E [\vert X_n \vert^2]\NN
\end{align}
because the variance of the sum of independent random variables equals the sum of their variances.
\end{remark}

Firstly, the averaged \magenta{desired signal} power\KJcancel{ of the desired signal} $\E_{\h,\CB}[\D_k]$ becomes
%
\begin{align}
\E_{\h,\CB}[\D_k]
&=\E_{\h,\CB}\big[\big\vert \tsuml_{m\in\A_k}\sqrt{\pmk} \h_{m,k}^\dg\w_{m,k}\big\vert^2\big]\NNL
&\stackrel{(a)}{=} \tsuml_{m\in\A_k}
\big\{
    \pmk\E_{\h,\CB}
    \big[
        \vert \h_{m,k}^\dg\w_{m,k}\vert^2
    \big]
\big\},\label{Aeqn:Appendix(a)}
\end{align}
where the equality $(a)$ holds from Remark 5 because of the following two reasons: 1) Each AP's beamforming is irrelevant to the beamforming of the other APs, making the random variables $\h_{m,k}^\dg\w_{m,k}$ for $m\in\A_k$ mutually independent; and 2) when UE $k$ is served by AP $m$ with ZF beamforming, UE $k$'s beamforming vector (i.e., $\w_{m,k}$) is independent of its own channel (i.e., $\h_{m,k}$) \cite{Lee2016, Jindal2006}, and it is satisfied that $\E[\h_{m,k}] = \zero_N$, which results in $\E[\h_{m,k}^\dg\w_{m,k}]=0$.

Also, on the RHS of \eqref{Aeqn:Appendix(a)}, we have
%
\begin{align}
&\E_{\h,\CB}
    \big[\big\vert \h_{m,k}^\dg\w_{m,k}\big\vert^2 ~\big\vert~ m\in\A_k \big]\NNL
&=\E_{\h,\CB}\big[\big\vert
    \sqrt{\beta_{m,k}N}
    \cdot
        \big(\tsuml_{l\in[L]}\sqrt{\kappa\mkl}
        g\mkl\a(\theta\mkl)^\dg\w_{m,k}
    \big)
        \big\vert^2\big]\NNL
&\stackrel{(b)}{=}\beta_{m,k}N\cdot
        \tsuml_{l\in[L]}
        \big\{
        \kappa\mkl
        \E_{\h,\CB}\big[\big\vert
        g\mkl\a(\theta\mkl)^\dg\w_{m,k}
        \big\vert^2\big]
        \big\}\NNL
&\stackrel{(c)}{\le}\beta_{m,k}N\cdot
        \tsuml_{l\in[L]}
        \big\{
        \kappa\mkl
        \E_{\h,\CB}
        \big[
            \vert g\mkl \vert^2
        \big]
        \big\}
\stackrel{(d)}{=} \beta_{m,k}N,\label{Aeqn:Appendix(d)}
\end{align}
%
where the equality $(b)$ holds from Remark 5 because at each UE, the dominant paths are independent of one another. Also, the inequality $(c)$ holds because $\vert \a( \theta\mkl) ^\dg \w_{m,k}\vert\le 1$, and the equality $(d)$ holds from $\E_{\h,\CB} [ \vert g\mkl \vert^2 ] = 1$ and $\sum_{l\in[L]} \kappa\mkl=1$. Thus,  plugging the inequality $\eqref{Aeqn:Appendix(d)}$ into $\eqref{Aeqn:Appendix(a)}$, we obtain $\E_{\h,\mathcal{C}}[\D_k] \le \D_k^{\downarrow}$.

\Blue

Secondly, the averaged SGI power $\E_{\h,\CB}[\I_k^\SGI]$ becomes
\begin{align}
    &\E_{\h,\CB}\big[\I_k^\SGI\big]
    =\tsuml_{\substack{j\in[K]    \\   j\ne k}}
    \E_{\h,\CB}
    \big[
        \big\vert\tsuml_{\substack{m\in\A_j   \\   m\in\A_k}}
        \sqrt{\pmk}
        \h_{m,k}^\dg\w_{m,j}\big\vert^2
    \big]\NNL
    &=
    \tsuml_{\substack{j\in[K]    \\   j\ne k}}
    \E_{\h,\CB}
    \big[\big\vert
        \tsuml_{\substack{m\in\A_j   \\   m\in\A_k}}
        \sqrt{\pmk}
        \big(\hh_{m,k}+\e_{m,k}\big)^\dg\w_{m,j}
    \big\vert^2\big]\NNL
    &\overset{(e)}{=}\tsuml_{\substack{j\in[K]    \\   j\ne k}}
    \E_{\h,\CB}\big[\big\vert\tsuml_{\substack{m\in\A_j   \\   m\in\A_k}}
    \sqrt{\pmk}
    \e_{m,k}^\dg\w_{m,j}\big\vert^2\big]\NNL
    &\overset{(f)}{=}\tsuml_{\substack{j\in[K]    \\   j\ne k}}
    \tsuml_{\substack{m\in\A_j   \\   m\in\A_k}}
    \big\{
    \pmk
    \E_{\h,\CB}
    \big[
        \vert \e_{m,k}^\dg \w_{m,j}\vert^2
    \big]\big\},\label{relaxedSGI1}
\end{align}
where the equality $(e)$ holds from the fact that in each UE group, the ZF beamforming vectors are obtained from the quantized channels, i.e., \eqref{practical_ZF_beamforming}. In addition, the equality $(f)$ holds from Remark 5 because of the following two reasons: 1) Each AP's beamforming is irrelevant to the beamforming of the other APs, making the random variables $\e_{m,k}^\dg\w_{m,k}$ for $m\in\A_k$ mutually independent; and 2) the decomposed error channel $\e_{m,k}$ is independent of $\w_{m,j}$ whenever $m\in\A_k$, and we have $\E[\e_{m,k}]=\zero_N$, which results in $\E[\e_{m,k}^\dg\w_{m,j}]=0$.

Also, on the RHS of \eqref{relaxedSGI1}, we have
\begin{align}
&\E_{\h,\CB}
    \big[\big\vert \e_{m,k}^\dg\w_{m,j}\big\vert^2 ~\big\vert~ m\in\A_k \big]\NNL
&=\E_{\h,\CB}\big[\big\vert
    \sqrt{\beta_{m,k}N}
    \cdot
        \big(\tsuml_{l\in[L]}\sqrt{\kappa\mkl}
        \delta\mkl\a(\theta\mkl)^\dg\w_{m,j}
        \big)
    \big\vert^2\big]\NNL
&\stackrel{(g)}{=}\beta_{m,k}N\cdot
        \tsuml_{l\in[L]}
        \big\{
        \kappa\mkl
        \E_{\h,\CB}\big[\big\vert
        \delta\mkl\a(\theta\mkl)^\dg\w_{m,j}
        \big\vert^2\big]
        \big\}\NNL
&\stackrel{(h)}{\le}\beta_{m,k}N\cdot
        \tsuml_{l\in[L]}
        \big\{
        \kappa\mkl
        \E_{\h,\CB}
        \big[
            \vert \delta\mkl \vert^2
        \big]
        \big\}\NNL
&\stackrel{(i)}{=}\beta_{m,k}N\cdot
        \tsuml_{l\in[L]}
        \big\{
        \kappa\mkl
        2^{-b\mkl}
        \big\},\label{eqn:long_eq24}
\end{align}
where the equality $(g)$ holds from Remark 5 because at each UE, the quantization errors of the dominant paths are independent of one another. Also, the inequality $(h)$ holds because $\vert \a( \theta\mkl) ^\dg \w_{m,j}\vert\le 1$, and the equality $(i)$ holds from Lemma 2. Thus, plugging the inequality \eqref{eqn:long_eq24} into \eqref{relaxedSGI1}, we obtain $\E_{\h,\CB} \big[\I_k^\SGI\big] \le\I_k^{\SGI\downarrow}$.

\Blue
Thirdly, the averaged IGI power $\E_{\h,\CB}[\I_k^\IGI]$ becomes
\begin{align}
\E_{\h,\CB}&\big[\I_k^\IGI\big]
=
    \tsuml_{\substack{j\in[K]   \\   j\ne k}}
    \E_{\h,\CB}\big[\big\vert\tsuml_{\substack{m\in\A_j   \\   m\notin\A_k}} \sqrt{\pmk} \h_{m,k}^\dg \w_{m,j}\big\vert^2\big]~\NNL
&\stackrel{(j)}{=}
    \tsuml_{\substack{j\in[K]   \\   j\ne k}}
    \tsuml_{\substack{m\in\A_j   \\   m\notin\A_k}}
    \big\{
    \pmk
    \E_{\h,\CB}
    \big[
        \vert \h_{m,k}^\dg \w_{m,j}\vert^2
    \big]\big\},\label{eqnA:appendix_IGI}
\end{align}
where the equality $(j)$ holds from Remark 5 because of the following two reasons: 1) Each AP's beamforming is irrelevant to the beamforming of the other APs, making the random variables $\h_{m,k}^\dg\w_{m,j}$ for $m\not \in\A_k$ mutually independent; and 2) each AP determines beamforming vectors solely by its associated UEs, independent of the channels of unassociated UEs, so $\h_{m,k}$ is independent of $\w_{m,j}$ whenever $m \not\in \A_k$, and we have $\E[\h_{m,k}] = \zero_N$, which results in $\E[\h_{m,k}^\dg\w_{m,j}]=0$.

Also, on the RHS of \eqref{eqnA:appendix_IGI}, we have
%
\begin{align}
&\E_{\h,\CB}
    \big[\big\vert \h_{m,k}^\dg\w_{m,j}\big\vert^2 ~\big\vert~ m\not\in\A_k \big]\NNL
&=\E_{\h,\CB}\big[\big\vert
    \sqrt{\beta_{m,k}N}
    \cdot
        \big(\tsuml_{l\in[L]}\sqrt{\kappa\mkl}
        g\mkl\a(\theta\mkl)^\dg\w_{m,j}
    \big)
        \big\vert^2\big]\NNL
&\stackrel{(k)}{=}\beta_{m,k}N\cdot
        \tsuml_{l\in[L]}
        \big\{
        \kappa\mkl
        \E_{\h,\CB}\big[\big\vert
        g\mkl\a(\theta\mkl)^\dg\w_{m,j}
        \big\vert^2\big]
        \big\}\NNL
&\stackrel{(l)}{\le}\beta_{m,k}N\cdot
        \tsuml_{l\in[L]}
        \big\{
        \kappa\mkl
        \E_{\h,\CB}
        \big[
            \vert g\mkl \vert^2
        \big]
        \big\}
\stackrel{(m)}{=} \beta_{m,k}N,\label{eqnA:appendix_IGI2}
\end{align}
where the equality $(k)$ holds from Remark 5 because at each UE, the dominant paths are independent of one another. Also, the inequality $(l)$ holds because $\vert \a( \theta\mkl) ^\dg \w_{m,j}\vert\le 1$, and the equality $(m)$ holds from $\E_{\h,\CB} [ \vert g\mkl \vert^2 ] = 1$ and $\sum_{l\in[L]} \kappa\mkl=1$. Thus,  plugging the inequality \eqref{eqnA:appendix_IGI2} into \eqref{eqnA:appendix_IGI}, we obtain $\E_{\h,\CB}[\I_k^\IGI]\le \I_k^{\IGI\downarrow}$.

\Black

\bibliographystyle{IEEEtran}

\end{document}